\documentclass[12pt,a4paper,final]{iopart}
\usepackage{iopams}
\usepackage{graphicx}
\usepackage[usenames,dvipsnames]{xcolor}
\usepackage{url,cite}
\usepackage{longtable}
\usepackage{tabularx,booktabs}

\usepackage[font=footnotesize,labelfont=bf]{caption}
\usepackage{multirow, bigstrut}
\graphicspath{ {./} }
\def\lsim{\mathop{\hbox{${\lower3.8pt\hbox{$<$}}\atop{\raise0.2pt\hbox{$\sim$}}
$}}} \def\gsim{\mathop{\hbox{${\lower3.8pt\hbox{$>$}}\atop{\raise0.2pt\hbox{$
\sim$}}$}}} 
 3
 
\renewcommand\fnsymbol[1]{{}^{\arabic{#1}}\relax}

\newcommand\eprint[1]{\href{http://arXiv.org/abs/#1}{#1}}

\providecommand\muK{\mu\mathrm{K}}
\definecolor{ballblue}{rgb}{0.0,0.3,.79}
\definecolor{dgreen}{rgb}{.1,.6,.1}
\definecolor{MyDarkRed}{rgb}{0.71,0.14,0.07}
\definecolor{MyDarkBlue}{rgb}{0.1,0,0.7}
\definecolor{charcoal}{rgb}{0.21, 0.27, 0.31}
\definecolor{goldenbrown}{rgb}{0.6, 0.4, 0.08}
\definecolor{lincolngreen}{rgb}{0.11, 0.35, 0.02}
\usepackage{hyperref}
\usepackage{array}

\usepackage{orcidlink}

\usepackage{geometry}
\usepackage{adjustbox}
\usepackage{esvect}
\usepackage[symbol]{footmisc}
\hypersetup{
    colorlinks=true,
    citecolor=blue,
    linkcolor=magenta,
    filecolor=cyan,      
    urlcolor=lincolngreen,
}

\begin{document}
\title[Temperature gradient probing the orientation of a finite Universe]{The variance of the CMB temperature gradient:
  \\ a dependence on the spatial orientation and anisotropy of a multiply connected Universe}

\author{Ralf Aurich$^{\,1}$\orcidlink{0000-0002-3206-9462}, Martin J France$^{\,2,*}$\orcidlink{0000-0002-2815-6686} and Frank Steiner$^{\,1}$\orcidlink{0000-0003-2039-0213}}

\smallskip

\address{$^1$Ulm University, Institute of Theoretical Physics, Albert-Einstein-Allee 11,
  \\$\ \ $D--89069 Ulm, Germany
  \\$^2$Universit\'e Lyon 1, ENS de Lyon, CNRS, CRAL, UMR 5574, Saint-Genis-Laval, France  
  \\$^*$Author to whom any correspondence should be addressed.

  \smallskip
  
  E-mail: martin.france@ens-lyon.fr, ralf.aurich@uni-ulm.de and frank.steiner@uni-ulm.de

  \bigskip
  
  }

\begin{abstract}
  In the present work we develop a model--independent method to reveal the underlying patterns of anisotropy and geometry in CMB temperature maps.
  This method, based on the $\rho-$statistics and a process of mask--sweeping, unveils faint but persistent geometrical and periodic structures characterising the
  anisotropy of the CMB in a universe with $3-$torus topology. We assess that the patterns detected over large ensembles
  of toroidal maps match geometrically to the layouts of expected circles-in-the-sky. The mask--sweeping method explores the whole CMB over 768 mask orientations.
  Thus we are able to re-verify the linear relation between $L$ the size of the cubic torus and $\rho$ the standard deviation of the normalised variance of the CMB
  temperature gradient. This confers a finite size $s$$\approx$3 Hubble lengths to our Universe with toroidal topology.
  Furthermore, it is shown that the violation of statistical isotropy is so faint that this has almost no influence on the linear relation between $\rho$ and $L$. 
  A topology like the cubic toroidal one imposes to our Universe periodic conditions (that can be seen as periodic boundary conditions)
  which constrain the global degree of inhomogeneity and a zero average curvature. 
  We demonstrate here that $\rho$ is not only a statistical signature of the spatial finiteness and size of our Universe (may its topology be simply or multiply connected)
  but also outlines the patterns of anisotropy of the CMB with the help of the mask--sweeping process.
\end{abstract}

\section{Introduction}
\label{sec:introduction}
Regarding questions about the spatial extent of our Universe such as, (i) can one decide if our Universe is spatially finite or infinite?,
(ii) is our Universe topologically simply connected or multiply connected? \cite{lumrouk1, luminet1, lumweeria1, aurich9, rouklum1},
(iii) do we live in a ``small Universe''? \cite{aurjanlusstei1, lehnquin1}, (iv) what is the spatial extent of our observable Universe and of the physical
universe? \cite{aluri1}, (v) what are the largest structures in our Universe?, cosmic topology (CT) brings an ensemble of
mathematical and geometrical concepts \cite{lachlum1, akrami1, copi1}. Indeed these questions are not constrained 
by the general theory of relativity (thereafter GR) since GR is a differential theory local by nature. Based on a homogeneous and isotropic solution to
the Einstein equation, the Friedmann-Lema\^itre-Robertson-Walker (FLRW) metric, the standard model of cosmology (thereafter $\Lambda$CDM model) is constructed

\smallskip

- accounting for most of the cosmological observations and measurements,

- having the Universe implicitly spatially infinite,

- with a matter--like constituent called cold dark matter (cold DM or CDM)

- and assuming a late accelerated expansion \cite{andriot1, michel1}.

\bigskip

The phenomenology of this acceleration is formalised in GR using a source term called dark energy (DE) characterised by a negative pressure $\textit{\small{P}}_{\Lambda}$
adding on the right hand side of the Einstein equation a source term $\Lambda \boldsymbol{g}_{\mu \nu}$ to the stress energy tensor $\boldsymbol{T}_{\mu \nu}$.
Alternatively the resolution can be made adding the term -$\Lambda \boldsymbol{g}_{\mu \nu}$ to the Einstein tensor $\boldsymbol{G}_{\mu \nu}$ on the left hand side.
$\Lambda$ is the cosmological constant and $\boldsymbol{g}_{\mu \nu}$ is the dynamical metric of space-time.
Distribution and dynamics of the cosmic matter and energy are idealized by an isotropic and homogeneous cosmic fluid. $\Lambda$ corresponds to a time and space invariant
negative equation of state w$_{\Lambda}$=$\textit{\small{P}}_{\Lambda}$/$\epsilon_{\Lambda}$= -1
($\textit{\small{P}}_{\Lambda}$ = pressure, $\epsilon_{\Lambda}$ = density of dark energy $\sim \Lambda$). 
Hitherto the physical nature of dark energy has not been identified while it can be noticed that the emerging of
DE coincides with the formation and growth, by accretion, of large-scale structures from inhomogeneities in the late
Universe at low redshift (z$\sim$0.5).

\smallskip

In a Universe with multiply connected topology such as the one of the cubic $3-$torus, under investigation here, there are periodic conditions that can be seen as
periodic boundary conditions even if the $3-$tori have no boundaries \cite{steiner1}. These periodic conditions impose stringent limits on the global
inhomogeneity and on the global spatial curvature, its growth and its magnitude \cite{buchert5, coley2}. It has to be reminded that the $3-$torus manifold, under examination here,
is inherently globally flat (non zero curvatures may appear locally but the average curvature is zero), globally homogeneous but globally anisotropic (see important definitions and
discussions about the $3-$torus in \cite{steiner1}).
A cubic $3-$torus has a side--length $s$=$L \cdot L_{H}$ (where $s$ and $L_{H}$=$c/H_{0}$, the Hubble radius, are in Mpc).

  The cubic 3-torus topology is a special case of the more general toroidal topology, which not only possesses different side-lengths,
  but the three side-lengths are not even perpendicular to each other, so that the fundamental cell is a parallelepiped (see e.g. \cite{steiner1}). We choose the cubic
  special case due to the strongly
  reduced computational effort in the calculation of the CMB sky maps. Because of the highly degenerated wave number spectrum, the computation of ensemble sizes of
  100\,000 members becomes feasible. 

\smallskip

At large spatial scales the CMB brings the most stringent constraints on the global topology and the overall acoustic properties of our
Universe. Large-scale temperature anisotropies of the CMB are dominated by the ordinary Sachs-Wolfe effect \cite{sachwolf1, sachwolfrep1}. CMB temperature
anisotropies $\Delta \textrm{T}$ at super-horizon scales ($l$$<$200) with adiabatic fluctuations are generated by the variations of the gravitational potential $\Phi_{\textrm{g}}$
predicted by the ordinary Sachs-Wolfe effect,
\begin{equation}
\label{SW}  
{\frac{\Delta \textrm{T}}{\textrm{T}}}=-\frac{1}{3} \Phi_{\textrm{g}}\;\;.
\end{equation}

\begin{figure}[htbp]
    \begin{center}
    \includegraphics[width=0.300\textwidth,angle=0]{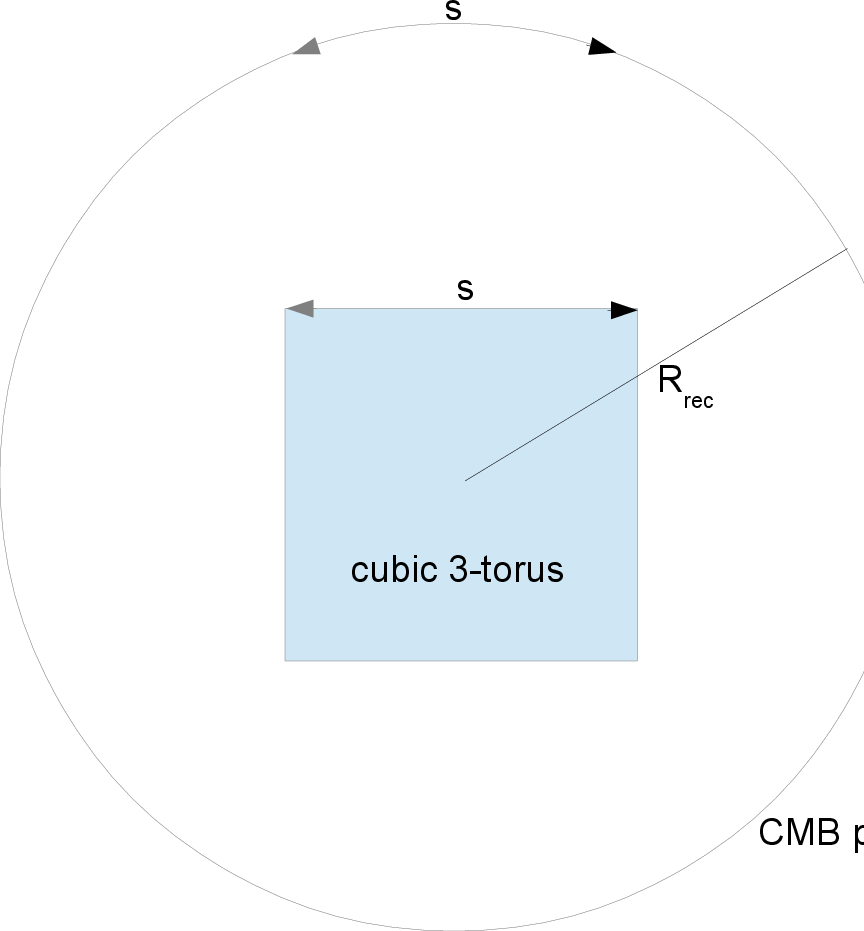}
    \caption{The cubic $3-$torus side--length $s$=$L~c/H_{0}$ projected 
      over the CMB fixes the cut-off at large wavelength
      $l_{\textrm{\scriptsize{cut-off}}}\simeq$2$\pi R_{\textrm{\scriptsize{rec}}}/s$.  
      (see table \ref{tab:table1}).}
    \label{fig:fig1}
    \end{center}
  \end{figure}

In a spatially flat and finite universe having a multiply connected topology, these fluctuations possess only a discrete spectrum of vibrational modes characterising for
instance, a cubic toroidal universe. In contrast to the infinite space the wave number spectrum ${\boldsymbol k}$ of the Laplacian in a toroidal universe is discrete and countably infinite, i.e.
there is no ultraviolet cut--off at large wave numbers. However, the wave numbers ${\boldsymbol k}$ are degenerate with very irregular multiplicities and with increasing mean value. The distinct
wave numbers of the cubic torus Universe with side--length $L$ are given by $k_{m}$=$\frac{2 \pi}{L} \sqrt{m}$ ($m \in \mathbb{N}$). Very importantly, there exists an infrared cut--off
$k_{1}(L)$=$|{\boldsymbol k}_{1}(L)|$=2$\pi/L$ since the fundamental mode (zero mode) $k_{0}$=0, corresponding to the monopole, is subtracted in the CMB. 
  Table \ref{tab:table1} and figure \ref{fig:fig1} illustrate how the infrared cut-off decreases with the side--length $s$=$L\frac{c}{H_{0}}$ of a
  cubic $3-$torus for a CMB radius $R_{\textrm{\scriptsize{rec}}}$. There is no such cut-off in a spatially infinite universe since
  s$\to$$\infty$ and $l_{\textrm{\scriptsize{cut-off}}}$$\to$0.

  \begin{table}[htbp]
    \begin{center}
      \begin{tabular}{cccc}
        \hline
        \multirow{2}{*}{$L$} & $s$ (Mpc~$h^{-1}$) & $l_{\textrm{\scriptsize{cut-off}}}$ & $\vartheta_{\textrm{\scriptsize{cut-off}}}(^{\circ})$ \bigstrut \\
        \cline{2-4} & $\frac{L ~ c}{H_{0}}$ & $\frac{2\pi R_{\textrm{\scriptsize{rec}}}}{s}$ & $\frac{360^{\circ}}{l_{\textrm{\scriptsize{cut-off}}}}$ \bigstrut \\
        \hline
        \multirow{1}{*}{0.5} & 2212.8 & 40.241 & 8.9 \bigstrut \\
        \hline
        \multirow{1}{*}{1.0} & 4425.6 & 20.121 & 17.9 \bigstrut \\
        \hline
        \multirow{1}{*}{1.5} & 6638.5 & 13.414 & 26.8 \bigstrut \\
        \hline
        \multirow{1}{*}{2.0} & 8851.3 & 10.060 & 35.8 \bigstrut \\
        \hline
        \multirow{1}{*}{3.0} & 13276.9 & 6.707 & 53.7 \bigstrut \\
        \hline
      \end{tabular}
      \caption{Table of the fundamental mode (associated to the wave number $k_{1}$=2$\pi$/$L$) 
        of the cubic $3-$torus. This table shows the corresponding infrared cut-off over the spherical surface 
        of projection of the CMB with respect to the $3-$torus side--length $s$ for a Hubble constant today 
        $H_{0}$$\simeq$67.74km~s$^{-1}$~Mpc$^{-1}$ and a distance to recombination $R_{\textrm{\scriptsize{rec}}}$$\simeq$14172.199Mpc~h$^{-1}$.
        The angle $\vartheta_{\textrm{\scriptsize{cut-off}}}$ provides an estimate of the infrared cut-off.}
      \label{tab:table1}
    \end{center}
  \end{table}

\begin{figure*}
    \centering
    \includegraphics[height=0.55\textheight,width=0.55\textwidth,angle=270]{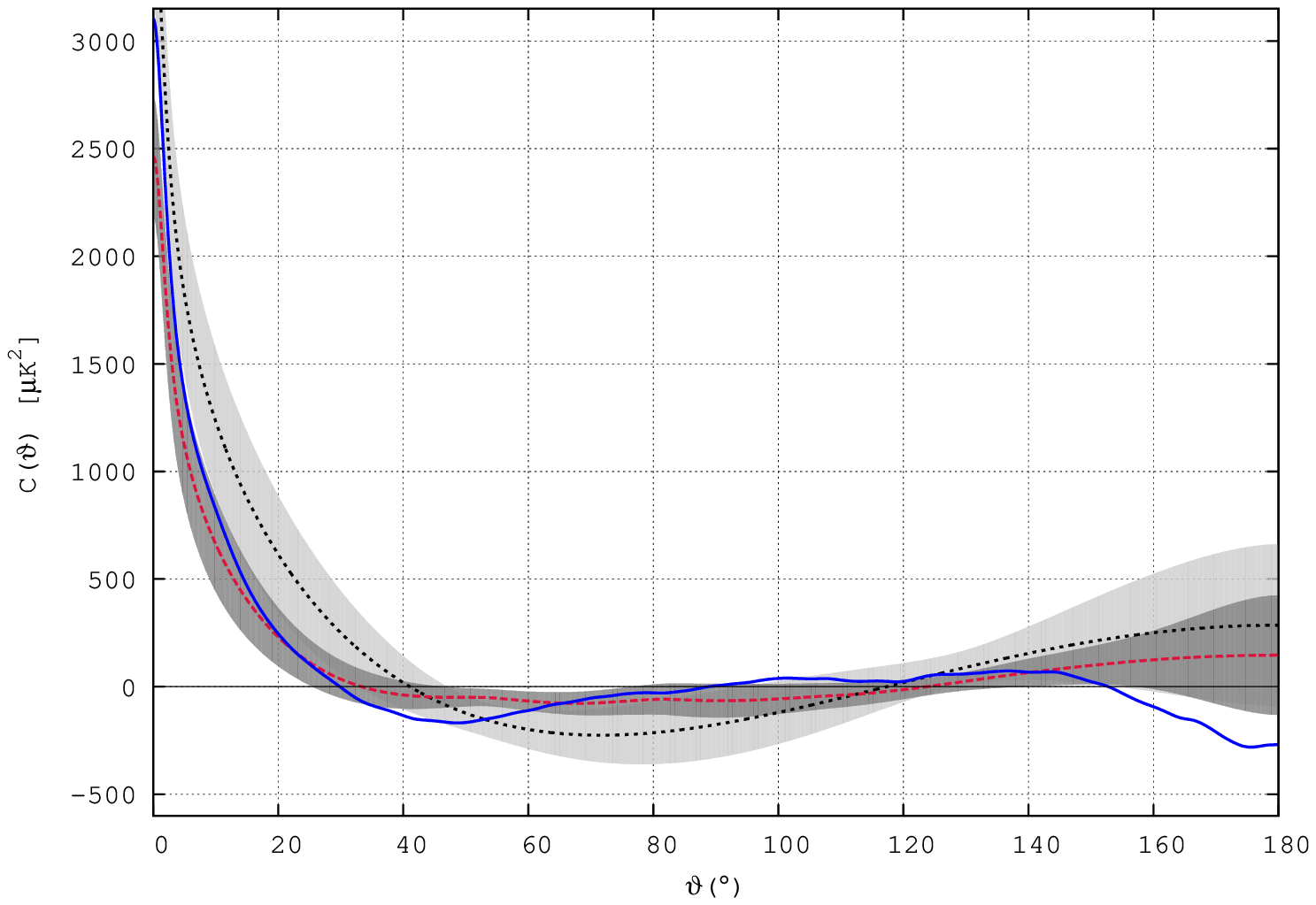}
    \caption{
      The average two-point correlation functions (2--pcf) are shown for ensembles consisting of $100\,000$ CMB maps computed from a full sky,
      where the parameters $N_{\hbox{\scriptsize side}}$=128, $l_{\hbox{\scriptsize max}}$=256 and $\hbox{FWHM}=120\hbox{arcmin}$ are used.
      The $\Lambda$CDM model ensemble is displayed as a small-dotted line in black and the torus ensemble at side--length $L=3.0$ as a large-dashed line in red.
      Furthermore, the $\pm 1\sigma$ regions are the shaded areas, dark for the torus at $L=3.0$, paler for the infinite $\Lambda$CDM model.
      The average 2--pcf of the observed CMB map (estimated by Planck collaboration through
      four different processing pipelines NILC, SEVEM, SMICA and Commander-ruler providing I intensity maps labelled I$\_$Stokes)
      is presented as a solid line (blue). The non-vanishing negative correlations of
      the blue curve beyond 150$^{\circ}$ are due to foreground galactic contamination. }
    \label{fig:fig2}
  \end{figure*}

  The angular 2-point correlation function (2-pcf) of the CMB is defined by 
  \begin{equation}
    \label{2pcf}
    C(\vartheta):=\left<\delta T({\boldsymbol{\hat n}})\delta T({\boldsymbol{\hat n'}}) 
    \right>\, \quad {\rm with} \quad {\boldsymbol{\hat n}}\cdot{\boldsymbol{\hat n'}} = \cos 
    \vartheta \;\;,
  \end{equation}
  where the brackets denote averaging over all directions $\boldsymbol{\hat n}$ and $\boldsymbol{\hat n'}$ that are separated by an angle $\vartheta$.
  Thus the 2-pcf of the spatially infinite $\Lambda$CDM model (see the black-dotted line in figure \ref{fig:fig2}) shows non-zero correlations up to 180$^{\circ}$.
  For a Hubble parameter value today of $H_{0}$$\simeq$67.74km~s$^{-1}$~Mpc$^{-1}$, the first angular acoustic peak of the power spectrum is at
  $\theta_{rec}$=0.0104077$\pm$0.0000032 radians and its spatial extent is $e_{rec}$=(147.5$\pm$0.6)Mpc giving in the $\Lambda$CDM model a distance to recombination 
  \begin{equation}
    {R_{\textrm{\scriptsize{rec}}}}=\frac{e_{rec}}{\theta_{rec}}\;\;,
  \end{equation}
  of $R_{\textrm{\scriptsize{rec}}}$=(14172.199$^{+62.027}_{-61.988}$)Mpc~h$^{-1}$.

  At large angular scales the 2-pcf of the CMB temperature anisotropies observed by ground-based microwave telescopes and space
  probes such as Planck vanishes for angles $>$ 60$^{\circ}$.
  Simulations of the CMB in the infinite $\Lambda$CDM model provide a 2-pcf which is clearly non-zero whatever the angle (excepted in the two intersecting points
  at $\sim$40$^{\circ}$ and $\sim$120$^{\circ}$).
  Thus the CMB of the $\Lambda$CDM model is
  in anomaly with the CMB observed at a statistical significance up to 1$\sigma$ in the 2-pcf in the angular range between 60 and 120$^{\circ}$
  (see figure \ref{fig:fig2}). For a detailed discussion and presentation of the cosmic topology complementary to \cite{lachlum1} we refer the reader to the work
  focusing on the CMB \cite{aurich9, aurich10}.

Simulation maps of the CMB in universe
models having a multiply connected topology (of characteristic sizes appropriately chosen) display a vanishing 2-pcf in the same wide angular range that
the observed CMB. Such models remain statistically consistent with the $\Lambda$CDM model for the CMB anisotropies at small spatial scales.  
The suppression of correlations at such large angles (i.e. for angles greater than 50$^{\circ}$ to 60$^{\circ}$) corresponds to the observed power asymmetry between
even
and odd low multipoles in the $C_{l}$ spectrum (see \cite{kim1}). The largest contribution would be expected from the dipole, which is usually considered as
purely dynamical and interpreted as the peculiar velocity of the observer with respect to the CMB rest frame. However, there are possibilities to separate
this effect \cite{aurich11}. 

\smallskip

The 2-pcf and the power spectrum, its counterpart in the $l$-domain, are main statistical tools allowing to investigate the
acoustical properties of the CMB. Furthermore the scalar quantity $\rho$ defined as the normalized standard deviation of the temperature
gradient field measured over the CMB is a signature of the size of the universe. Thus our paper of 2021
\cite{aurich10} investigates the evolution of $\rho$ as a function of the spatial size of the Universe and demonstrates that there is a hierarchical
and linear relation between the size of the universe (universe model and observable Universe) and the average value taken by $\rho$ over the CMB.
The multiply connected manifold considered for
the finite model is the cubic 3-torus with various side--lengths $L$. The $\rho$ value of the CMB observed by Planck leads through the linear law
$L$=$a\langle\rho\rangle$+$b$ to a toroidal size of our Universe of about $L$=3. 

\smallskip

A finite universe raises the question to know how it is spatially oriented and studies have been endeavored to analyse the spatial-correlations in 
a toroidal universe and to compare to correlations in our observable Universe \cite{roukema1,roukema2, aurich12,aurich13}. These studies do not 
reject or confirm the toroidal topology and its privileged orientation since e.g. \cite{aurich12,aurich13} points at a non--negligible positive rate of the 
spatial-correlation method.

We would like to note that anomalies like the hemispherical power asymmetry in the CMB will not be explained by the cubic 3-torus topology because too
many copies of the fundamental cell are evenly distributed over the surface of last scattering. To explain such anomalies with the 3-torus being
homogeneous, one would need more asymmetrical fundamental cells.  

\smallskip

In a universe having finite spatial sections such as the cubic 3-torus it is expected that not only the 2-pcf shows a 
cut-off at the largest angular scales but also that geometrical patterns of identified circles-in-the-sky (CITS) should form, differently according 
to the size and the proportion of the torus. However such CITS are not observed over the CMB temperature map. A hypothesis is that the CITS signal is
of lesser amplitude than expected and is drowning in other temperature fluctuations. Indeed, the expected 
CITS patterns appear in CMB simulation maps having for instance the multiply connected topology of the cubic 3-torus.
Our study \cite{aurich10} suggests already that a too high amplitude of the late-time ISW could make the CITS geometric signal not detectable on the CMB.
The threshold for the CITS detection due to false positive cases is based on the $\Lambda$CDM model. But figure \ref{fig:fig2} reveals that the 2-pcf of the
  Planck data is significantly lower than that of the $\Lambda$CDM model even for small angles $< 40^\circ$ which are relevant for the determination of the threshold.
  This points to an inconsistency in the determination of the CITS threshold. 
Among the physical processes contributing to the CMB temperature anisotropies we observe from Earth, the integrated Sachs-Wolfe effect (ISW) is the 
change of frequency a CMB photon experiences in its propagation to our detectors as a consequence of the gradients of gravitational potentials 
encountered during the expansion of large-scale structures. As a consequence of this gravitational redshift, in the adiabatically expanding
$\Lambda$CDM model, CMB photons lose energy $\Delta_{E}$=h($\nu_{0}$-$\nu_{CMB})$$<$0 and temperature
$\Delta T_{ISW}$=$T_{0}$-$T_{CMB}$$<$0 until they arrive at $z$=0. ISW is a direct signature of spacetime dynamics.
Therefore ISW allows to test different scenarios of accelerated expansion and the possible redshift-wise evolution of DE \cite{seraille1}. 
An ISW ratio A$_{ISW}$=$\Delta T_{obs}/\Delta T_{\Lambda CDM}$=1 means conformity of observations to
the $\Lambda$CDM model. However, results obtained over the interval 2008-2019 \cite{granett1,kovacs1,kovacs2} of investigations report an excess of late-time
ISW from SDSS, BOSS and DES Y3 data.
An ISW in excess A$_{ISW}$=5.2$\pm$1.6 (in the redshift range 0.2$<$z$<$0.9) is found when both DES and BOSS data are used.
A sign change of ISW at very small redshift is recently found by \cite{hansen1}. Going against the predictions for the local ISW, CMB photons traversing voids
are much hotter than predicted while they are much colder than expected passing through the overdense regions of galactic halos.
An ISW in excess could hint at a possible explanation of contamination of the CITS signal not predicted in current simulations of the $\Lambda$CDM model.
It has to be seen how the discovery of redshift-evolving DE (see e.g. \cite{abdulkarim1} and \cite{lodha1}) could impact the estimation of the ISW
signal and its amplitude at low redshift.
Thus, the absence of observation of CITS cannot exclude the possible existence and detection of a multiply-connected universe through other methods,
such as statistical ones. Furthermore, recently, two investigations in CMB analysis provide other evidences that our Universe could be spatially finite. (i)
$\rho$ the variance of CMB temperature gradient displays a monotonic ordering of the universe models as a function of their spatial sizes. The larger the
universe model the smaller $\rho$ is \cite{aurich10}. (ii) A hierarchical ordering, monotonic, appears also through the analysis of CMB model maps with
toroidal topology using the Betti-functionals \cite{aurich14} and the concept of persistence \cite{pranav3, aluri1}. 
Moreover homology \cite{pranav10} detects a significant breakdown of the isotropy expected by the $\Lambda$CDM model in the map of CMB temperature observed by Planck.
It has to be verified if this deviation from anisotropy is consistent with the global anisotropy intrinsic to the $3-$torus multiply connected topology.

\smallskip

  In this paper we address the question for the violation of statistical isotropy only in the case of the intensity maps. Polarization maps have the advantage
  that there is no analogue to the integrated Sachs-Wolfe effect, so that a large contribution to the CMB is absent, which modifies the structures on the CMB.
  However, the polarization is much harder to measure, but see \cite{samandar1}. 

The $\Lambda$CDM model in an infinite Universe is thus statistically challenged at very different confidence levels provided by various precisions observations of the CMB
being in anomaly, in discrepancy or in tension with this model and at very different spatial scales and using different methods.

Further investigations on the sensitivity of the quantity $\rho$ to the acoustic modes at large-scale are presented in the present paper which is structured as follows.
In section \ref{sec:methodology} we describe globally the methodology of mask--sweeping adopted to detect the spatial orientation of the $3-$torus
over CMB masked maps using the normalized standard deviation of the temperature gradient field $\rho$ and its two components $\sigma_{0}$ and $\sigma_{1}$.
The way the $3-$tori are spatially oriented by construction is presented and the conventions used are described to explore the
different orientations of the mask based on Euler's rotation theorem. We present the three different kinds of CMB map ensembles analysed here and in
our previous study on $\rho$ \cite{aurich10}. These CMB maps correspond to the infinite $\Lambda$CDM model, the finite multiply connected model of the
$3-$torus and the unique CMB map observed by Planck processed through 5 different pipelines. Section \ref{sec:signature} shows the central result of this work
that the 3-dimensional spatial orientation of the $3-$torus is revealed in the plots by a 2-dimensional periodic distribution of spots while no such features appear
for the infinite $\Lambda$CDM model. In section \ref{sec:linear} the mask--sweeping ansatz confirms the linear relation existing between $\rho$ and $L$ the size of the fundamental
cells as demonstrated in \cite{aurich10}. 

\smallskip

\section{Principle of the method of mask sweeping}
\label{sec:methodology}

Large ensembles of CMB temperature maps, (i) in the infinite universe of the $\Lambda$CDM model and (ii) in a $\Lambda$CDM model with cubic $3-$toroidal topology have
nearly the same high level of isotropy and homogeneity. However, at angular scales larger than 50$^{\circ}$ the 2 models differ since the $2-$point correlation
function of the toroidal universe (cubic of side--length $L$=3.0) shows more closely the same absence of correlations than the observed CMB but, in the contrary, the
infinite $\Lambda$CDM model presents non zero correlations whatever the separation angle \cite{aurjanlusstei1, aurich10}.
Thus this work provides a statistical evidence that $\rho$ allows to sort hierarchically the universe models in function of their size.

\smallskip

The present work resorts to the known orientation of the cubic tori. By construction, all the CMB $3-$torus maps have their cubic torus cells spatially
oriented exactly in the same way with respect to the galactic system of coordinates.

\begin{figure}[!htb]
    \includegraphics[height=0.35\textheight,width=1.00\textwidth, angle=0]{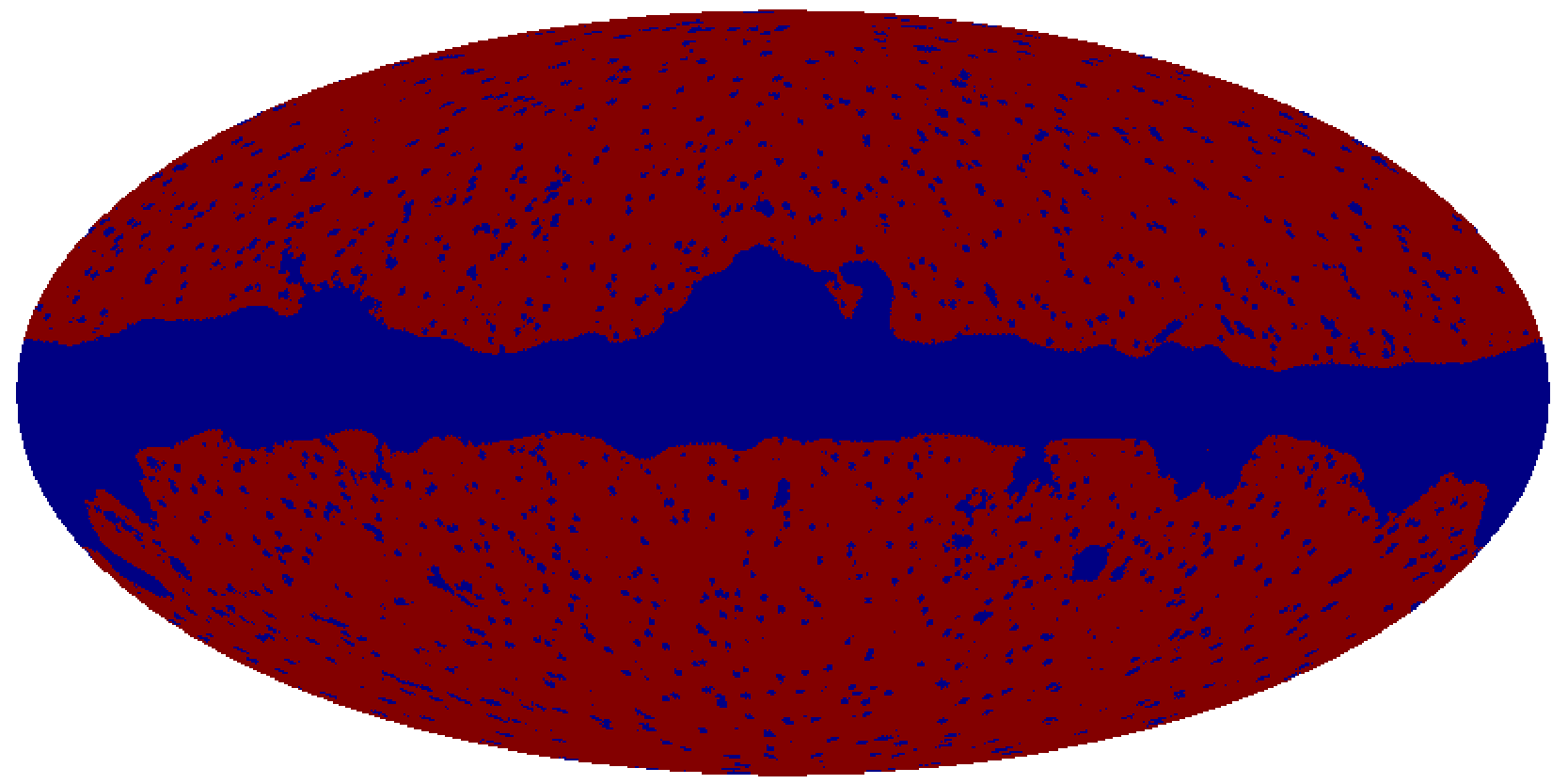}
\caption{This figure shows the Mollweide projection of the U73 foreground mask provided by the Planck collaboration at HEALPix resolution $N_{side}$=128.
  This mask leaves $\sim$73$\%$ of the CMB visible (regions in red). This is the mask used in the mask--sweeping process. For 768 orientations of this mask,
  the average quantities $\left<\sigma_{0}\right>$,  $\left<\sigma_{1}\right>$ and $\left<\rho\right>$ of the ensembles of CMB temperature maps are calculated. The
  different orientations of this mask are calculated using the HEALPix routine rotate${\_}$alm applied to the Euler angles $\Phi_{or}$, $\Theta_{or}$ while 
  the third Euler angle is left constant at zero (see \ref{app:euler} for details). 
}
\label{fig:fig3}
\end{figure}

The method of detection is the following: for the subset of $n_{\textrm{\scriptsize m}}$ realisation maps in the selected model universe, the average values of the
quantities $\sigma_{0}$, $\sigma_{1}$ and $\rho$ are calculated for each of the $n_{\textrm{\scriptsize or}}$ different orientations of the foreground mask (U73) shown 
in figure \ref{fig:fig3}. These $n_{\textrm{\scriptsize or}}$ different orientations of the mask over the CMB have to be evenly distributed on the $2-$sphere.
This is the case in the HEALPix tesselation where CMB pixels are of equal area. The spherical coordinates $\Phi_{\textrm{\scriptsize or}}$ (galactic longitude)
and $\Theta_{\textrm{\scriptsize or}}$ co-latitude of the center of each HEALPix pixel are used to rotate the foreground mask over the ensemble of n$_{\textrm{\scriptsize m}}$ CMB
maps in $n_{\textrm{\scriptsize or}}$ orientations. An HEALPix resolution $N^{or}_{side}$ allows to get the $n_{\textrm{\scriptsize or}}$=12($N^{or}_{side}$)$^{2}$ needed different
orientations of the foreground mask. In the present study it is the foreground mask U73 (source file COM$\_$MASK$\_$CMB-union$\_$2048$\_$R1.10.fits)
  \cite{planck1} which is rotated over the 768 different orientations allowed by the HEALPix tesselation at $N^{or}_{side}$=8 (see \ref{app:euler} for details on our
implementation of Euler angles on the $2-$sphere).

\bigskip

For each orientation
$\boldsymbol{\vartheta_{\textrm{\scriptsize or}}}$ = $\boldsymbol{\vartheta_{\textrm{\scriptsize or}}}$($\Phi_{\textrm{\scriptsize or}}$, $\Theta_{\textrm{\scriptsize or}}$)
of the mask, the average values of three statistical quantities measured over the $n_{\textrm{\scriptsize m}}$ masked CMB temperature maps, are calculated:

\smallskip

(i) the standard deviation
\begin{equation}
\label{equ:s0aver}
\left<\sigma_{0}(\boldsymbol{\vartheta_{\textrm{\scriptsize or}}})\right>\;\;,
\end{equation}

\smallskip

(ii) the standard deviation of the variance of the local temperature gradient
\begin{equation}
\label{equ:s1aver}
\left<\sigma_{1}(\boldsymbol{\vartheta_{\textrm{\scriptsize or}}})\right>\;\;,
\end{equation}

\smallskip

(iii) and the normalized standard deviation of the gradient field $\boldsymbol{G}$ of temperature anisotropy
\begin{equation}
\label{equ:rhoaver}
\left<\rho(\boldsymbol{\vartheta_{\textrm{\scriptsize or}}})\right>\;\;.
\end{equation}

\smallskip

The reader may refer to the complete definitions in \cite{aurich10}.
$\rho$ is calculated using the spherical coordinates of longitude $\phi$ and co-latitude $\theta$ over a CMB temperature map

\begin{equation}
\label{equ:rho}
{\rho}:= \sqrt{ \frac{\left<G_{\theta}^2+G_{\phi}^2\right>}{\sigma_{0}^{2}} } = \sqrt{\frac{\sigma_{1}^{2}}{\sigma_{0}^{2}}}\;\;.
\end{equation}

\bigskip

The CMB maps investigated in this study use the resolution $N_{side}$=128 (196608 HEALPix pixels of diagonal 27'5 and side 19'4), $l_{max}$=256 and
a Gaussian smoothing of $\vartheta_{G}$ = $2^{\circ}$. We use $100\,000$ simulation maps of the CMB in the infinite $\Lambda$CDM model
(according to the cosmological parameters of Planck 2015 \cite{planck2}). Furthermore, $100\,000$ maps of the CMB in a $\Lambda$CDM model with cubic $3-$torus
topology are generated for each of the five side--lengths $L$=0.5, 1.0, 1.5, 2.0 or 3.0 where $L$=$s/L_{H}$.
The unique observable CMB radiation map is processed through three different pipelines devised by the Planck
collaboration to reduce the foreground contamination. These pipeline maps are NILC, SMICA and SMICA-no SZ (see e.g. \cite{tianyue1, remazeilles1}). The
contaminated or obfuscated CMB regions due to our Milky Way, numerous extragalactic sources and gas clouds are reconstructed by Planck collaboration using the
inpainting method applied to each pipeline (Inpainted I intensity Planck maps) \footnote{One should mention that \cite{hamann1} devising and using an original mathematical
tool (direction-dependent CMB probe) detects clear departure from Gaussianity and isotropy in all the inpainted $Planck$ maps.}. 

\section{Signature of the spatial orientation of the three-torus with $\rho$}
\label{sec:signature}

\begin{figure}[!htb]
  \includegraphics[height=0.38\textheight,width=1.1\textwidth, angle=0]{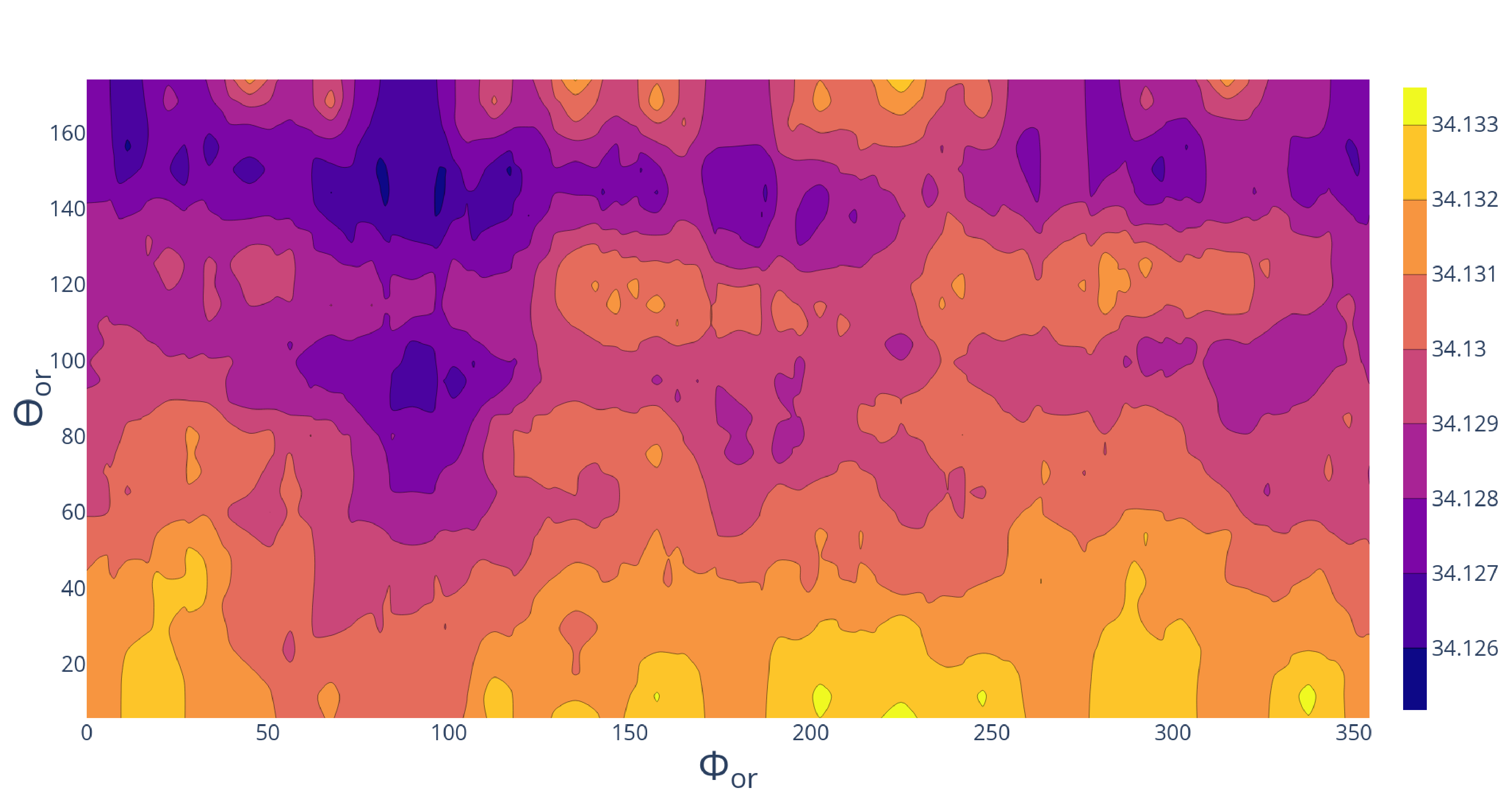}
    \caption{This figure shows in equirectangular projection (longitude $\Phi_{or}$ on the horizontal axis and co-latitude $\Theta_{or}$ on the vertical axis) the CMB $\rho$-carpet of
      the infinite $\Lambda$CDM model calculated by 768 rotations of the U73 mask over an ensemble of 100$\,$000 maps. $\rho$ values are given in the vertical bar.
    }
\label{fig:fig4}
\end{figure}

The main result of this paper is presented in 2-dimensional equirectangular projection, of coordinates $\Phi_{or}$ (longitude) and $\Theta_{or}$ (co-latitude), by graphical
representations ($\rho-$carpets, $\sigma_{0}-$carpets or $\sigma_{1}-$carpets). The principle based on the rotation of the CMB foreground mask (see figure \ref{fig:fig3}) relative
to the CMB temperature maps is described in detail in chapter \ref{sec:methodology}. The analysis is based on the
calculation of 3 different statistical quantities averaged over 100$\,$000 maps for each of the 768 mask orientations in each model.
Figure \ref{fig:fig4} shows the $\rho-$carpet of the CMB in the infinite $\Lambda$CDM model. This figure displays no specific periodic pattern whatever the size of the ensemble of maps
chosen. The variation $\Delta\rho$ of $\rho$ is of the order of 0.007 and thus much smaller than those found in the cubic $3-$torus topologies as demonstrated in the following
  five figures. Since the infinite $\Lambda$CDM model is statistically isotropic, this gives the order of anisotropy left by averaging over 100$\,$000 models. 

\smallskip

\begin{figure}[!htb]
\includegraphics[height=0.38\textheight,width=1.10\textwidth, angle=0]{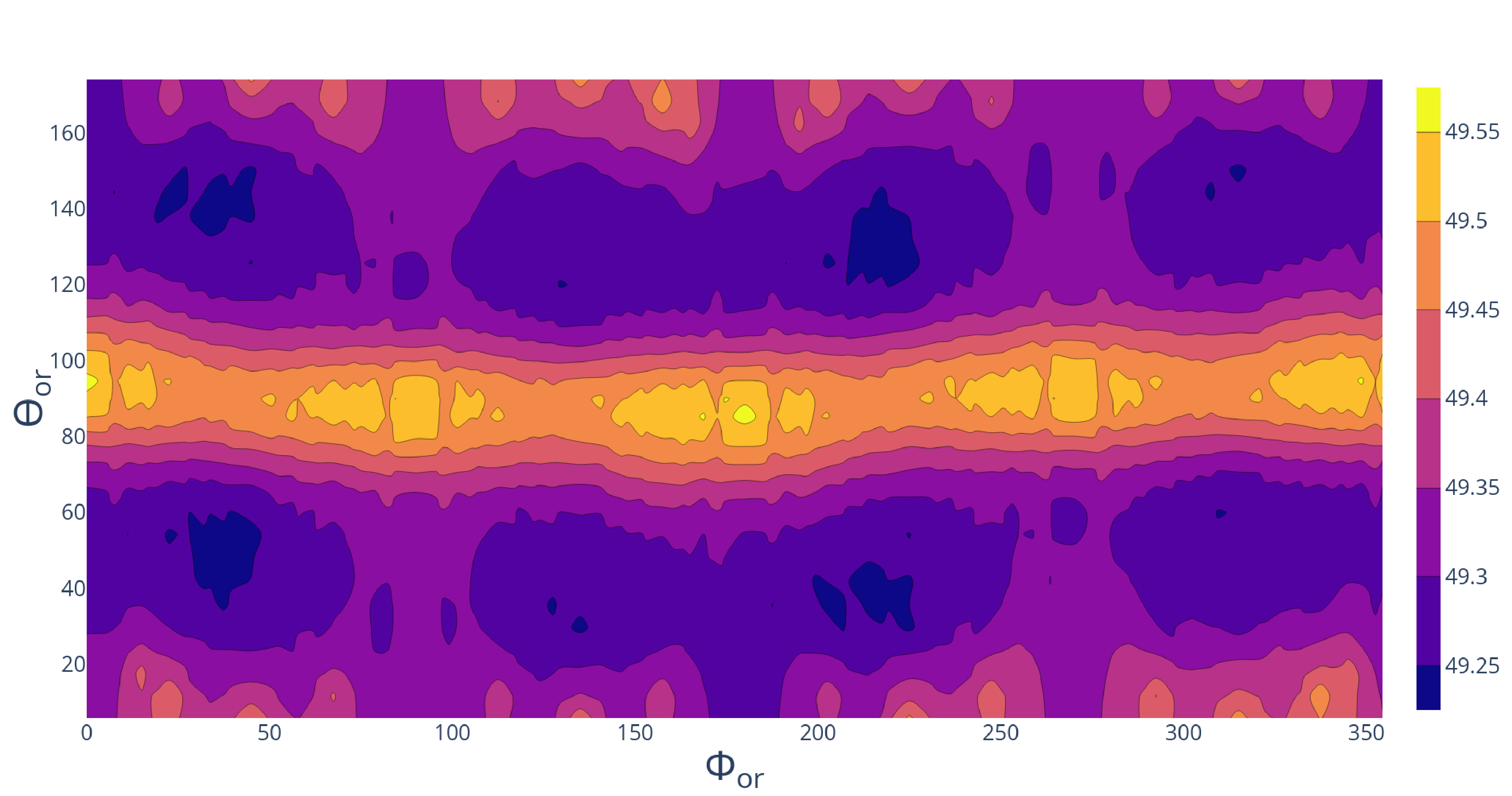}
\caption{This figure shows the CMB $\rho$-carpet for the $\Lambda$CDM model with cubic toroidal topology at $L$=0.5 calculated by 768 rotations of the U73 mask over 100$\,$000 maps.}
\label{fig:fig5}
\end{figure}

\begin{figure}[!htb]
\includegraphics[height=0.38\textheight,width=1.10\textwidth, angle=0]{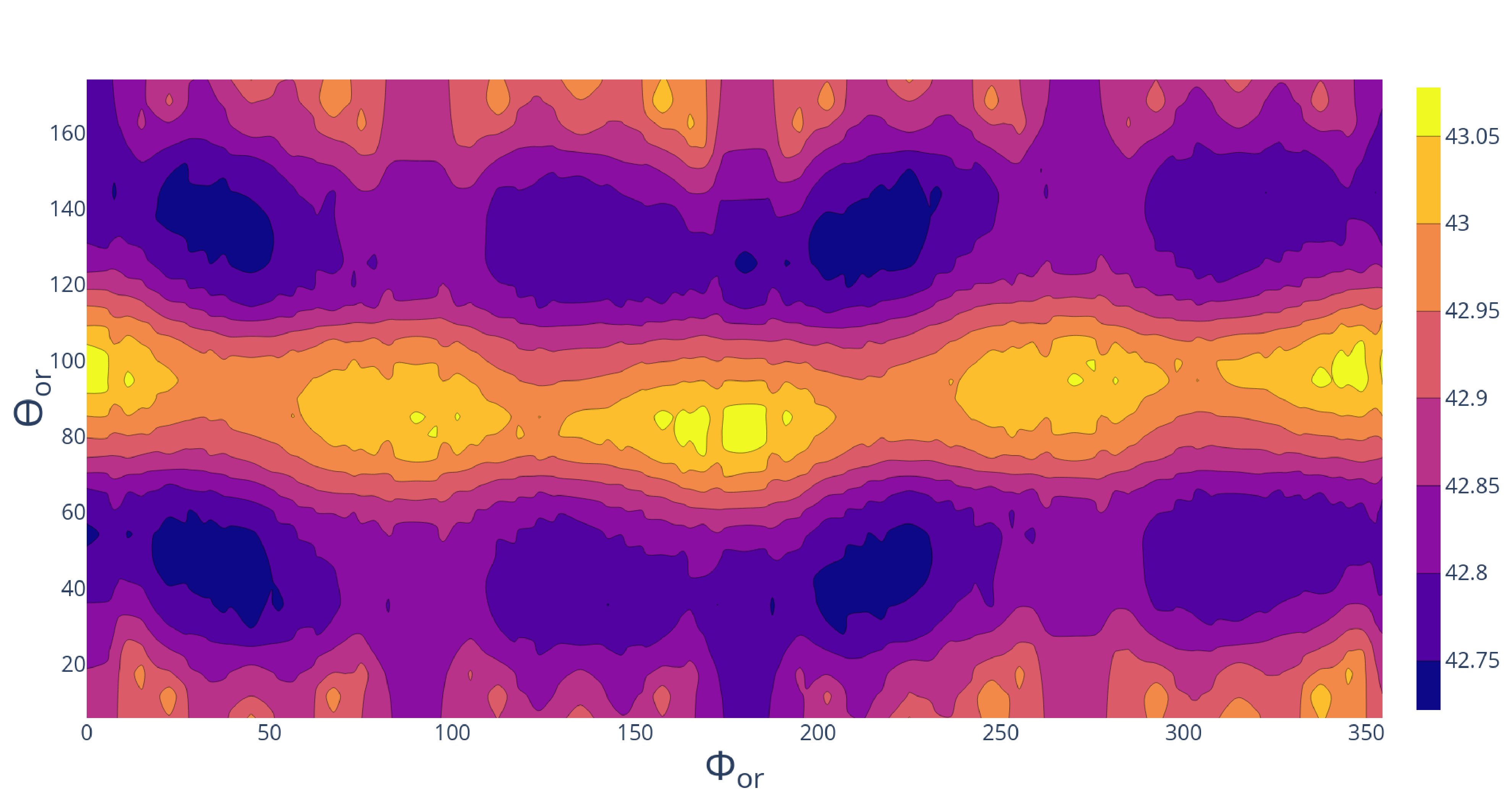}
\caption{This figure shows the CMB $\rho$-carpet for the $\Lambda$CDM model with cubic toroidal topology at $L$=1.0 calculated by 768 rotations of the U73 mask over 100$\,$000 maps.}
\label{fig:fig6}
\end{figure}

\begin{figure}[!htb]
\includegraphics[height=0.38\textheight,width=1.10\textwidth, angle=0]{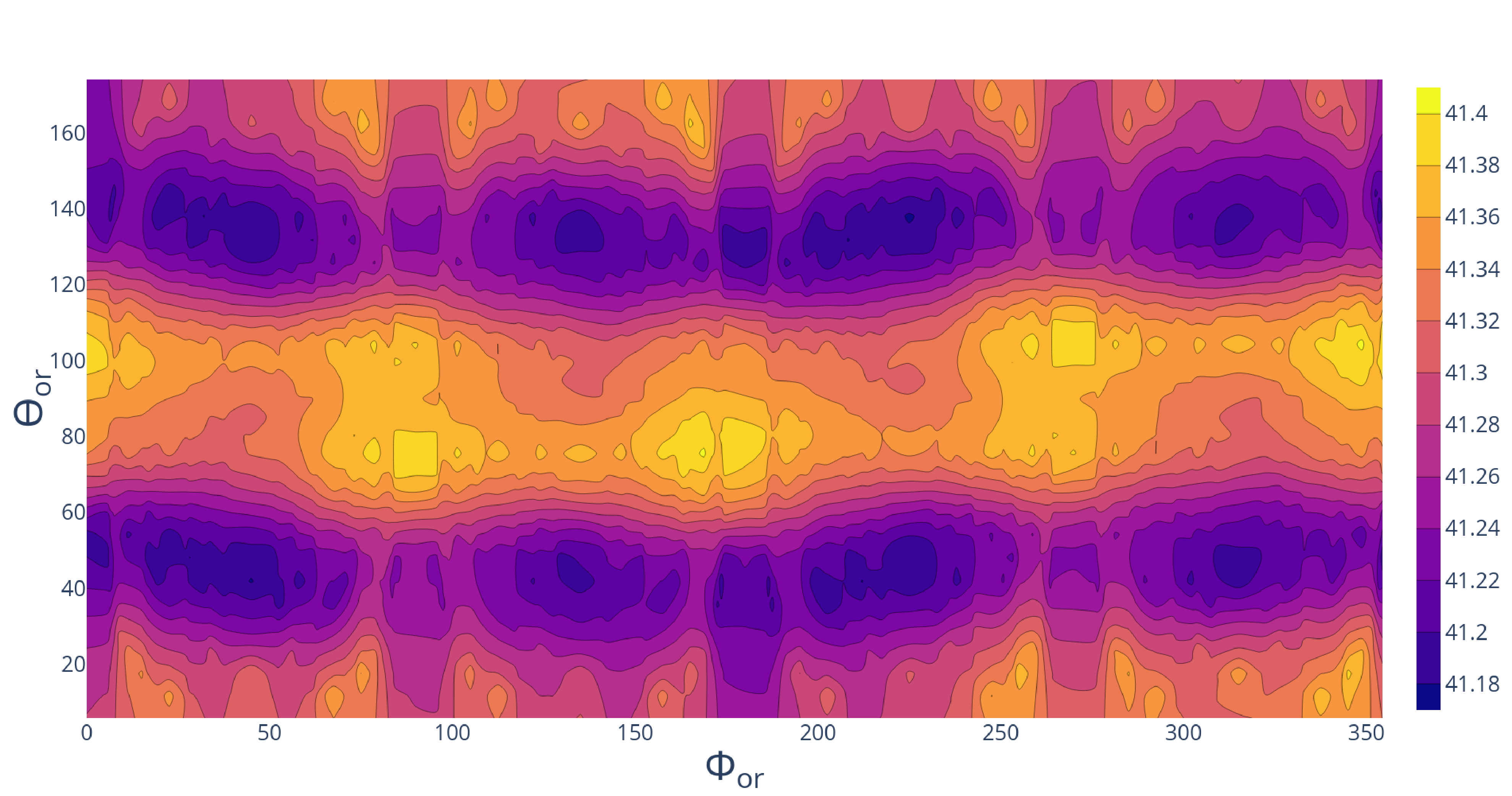}
\caption{This figure shows the CMB $\rho$-carpet for the $\Lambda$CDM model with cubic toroidal topology at $L$=1.5 calculated by 768 rotations of the U73 mask over 100$\,$000 maps.}
\label{fig:fig7}
\end{figure}

\begin{figure}[!htb]
\includegraphics[height=0.38\textheight,width=1.10\textwidth, angle=0]{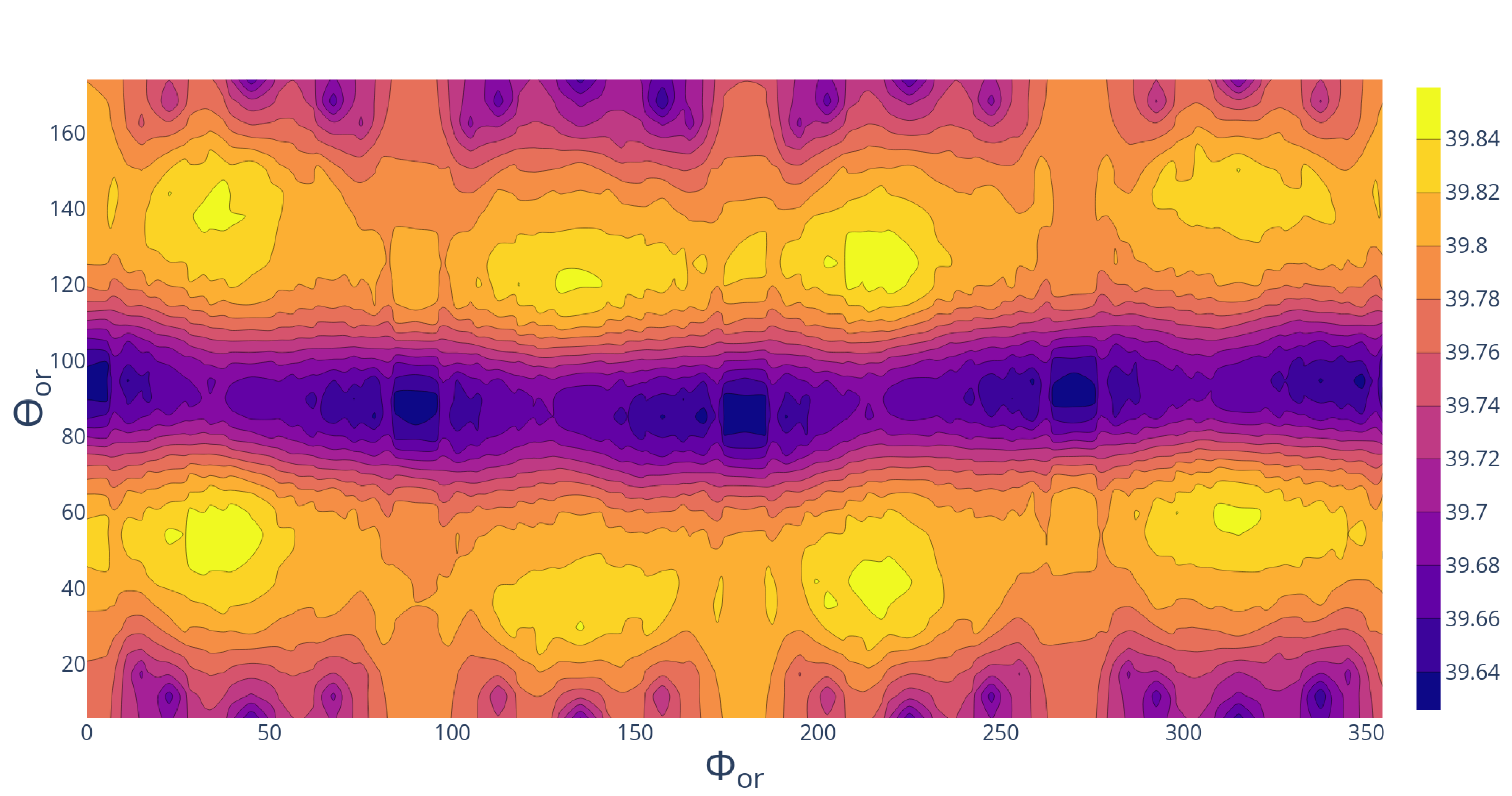}
\caption{This figure shows the CMB $\rho$-carpet for the $\Lambda$CDM model with cubic toroidal topology at $L$=2.0 calculated by 768 rotations of the U73 mask over 100$\,$000 maps.}
\label{fig:fig8}
\end{figure}

\begin{figure}[!htb]
\includegraphics[height=0.38\textheight,width=1.10\textwidth, angle=0]{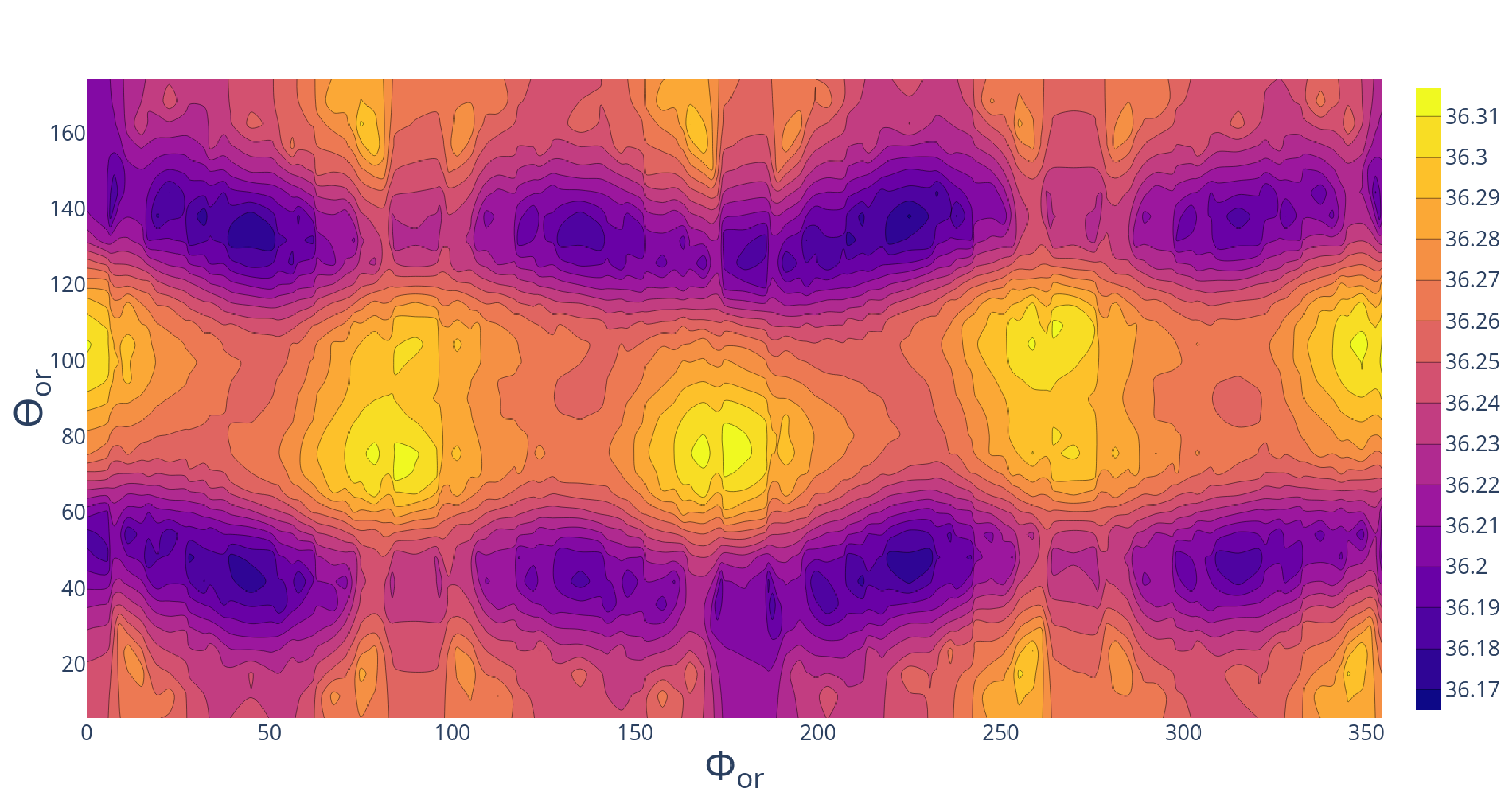}
\caption{This figure shows the CMB $\rho$-carpet for the $\Lambda$CDM universe model with cubic toroidal topology at $L$=3.0 calculated by 768 rotations of the U73 mask over 100$\,$000 maps.}
\label{fig:fig9}
\end{figure}

In the contrary to figure \ref{fig:fig4} which shows no specific and repetitive pattern, the five following $\rho-$carpets of the CMB with cubic $3-$toroidal topology in figures
\ref{fig:fig5}, \ref{fig:fig6}, \ref{fig:fig7}, \ref{fig:fig8} and \ref{fig:fig9} do present a periodic pattern layout of alternate bright and dark spots regularly distributed.

Figures \ref{fig:fig5}, \ref{fig:fig6}, \ref{fig:fig7}, \ref{fig:fig8} and \ref{fig:fig9} show the $\rho$-carpets in equirectangular projection (longitude $\Phi_{or}$ and co-latitude
$\Theta_{or}$) of the CMB in a $\Lambda$CDM model with finite cubic $3-$torus topology of side--lengths $L$=0.5, 1.0, 1.5, 2.0 or 3.0.
One observes structures of the order $\Delta\rho\simeq 0.25$ being significantly larger than the expected level due to the finite ensemble size of 100$\,$000 models.
  Nevertheless, the variations are small compared to the $\rho$ values of the different torus sizes $L$. 

\begin{figure}[!htb]
\includegraphics[height=0.38\textheight,width=1.10\textwidth, angle=0]{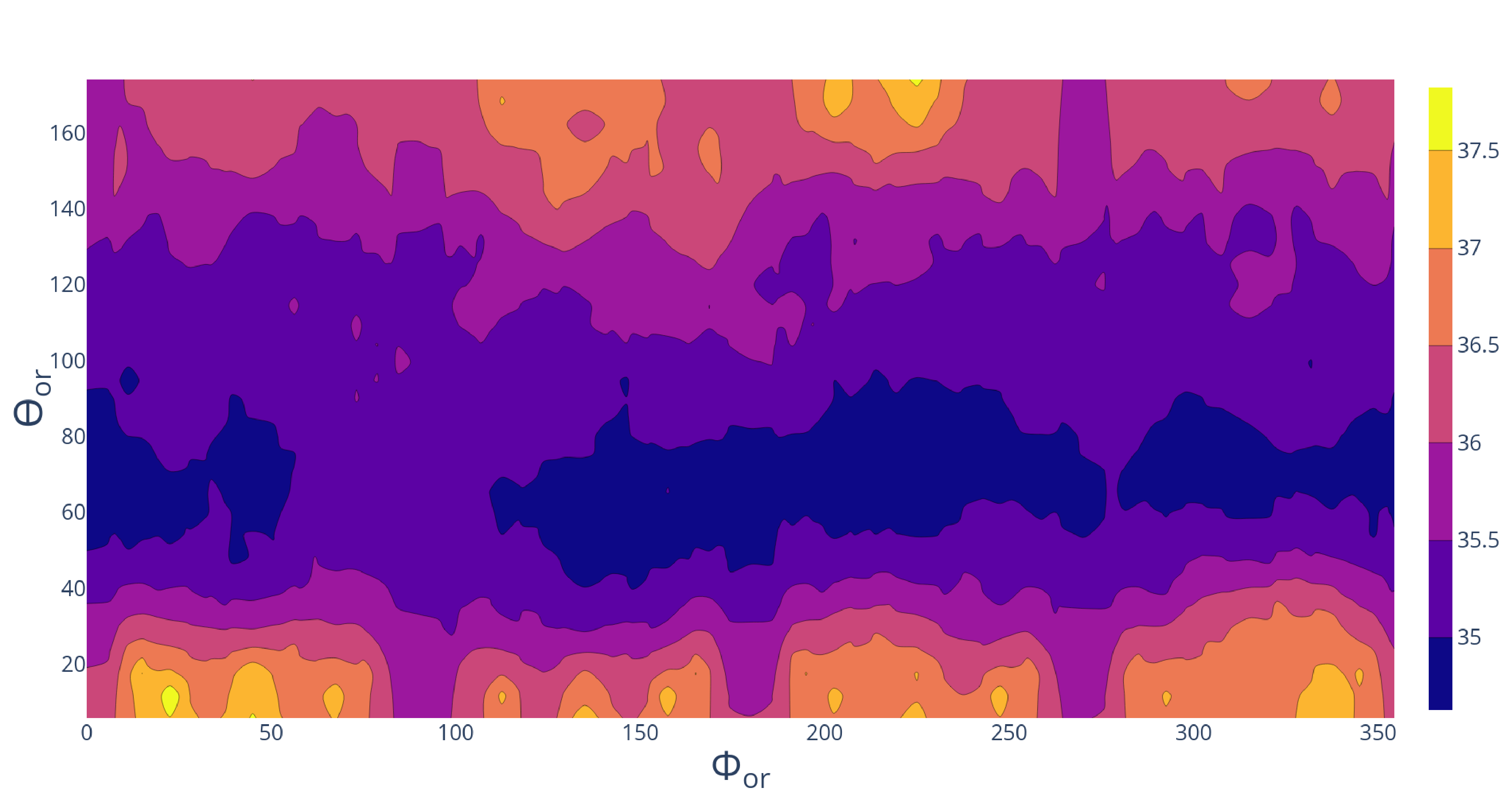}
\caption{This figure shows the $\rho$-carpet of the unique CMB (observed by Planck) and processed through the three pipelines NILC, SMICA and SMICA no SZ. 
  This $\rho$-carpet is calculated by 768 rotations of the U73 mask over the 3 by-product maps.}
\label{fig:fig10}
\end{figure}

Figure \ref{fig:fig10} shows the $\rho-$carpet of the unique CMB radiation map observed by Planck. The averaging is made over three by-product maps provided
through the processing pipelines NILC, SMICA and SMICA corrected from the Sunyaev-Zeldovich effect. Using our mask--sweeping method the signature of orientation begins to appear
vaguely for an ensemble of one hundred maps with $3-$torus topology. Notice also that torus maps and maps of the infinite $\Lambda$CDM model are deprived over the entire surface of foreground
noise due to the Milky Way and extragalactic sources. By comparison the Planck by-product maps would be valid only if the mask is
oriented precisely over the galaxy and the numerous sources of extragalactic contamination. Thus 768 - 1 orientations of the mask unveil artificially recovered regions of the real microwave sky.
These regions are processed in order to reconstruct as much as possible the assumed statistical properties of the invisible or contaminated CMB (Inpainted I intensity Planck
maps labelled I$\_$Stokes$\_$Inp). In the case of the U73 mask 27$\%$ of the sky is not taken into account. Regarding the issue of the CMB mask the reader may refer to the
study ``Can one reconstruct the masked cosmic microwave background sky?'' in \cite{aurich15}.

\clearpage
\newpage

\begin{footnotesize}
  \begin{center}
    \begin{longtable}[!htb]{@{}llllllllll@{}}
      \hline
      $L$ & $\langle\sigma_{0}\rangle$ & $\langle\sigma_{1}\rangle$ & $\langle\rho\rangle$ & $median$ & $\delta_{1}$ & $\Sigma$ & $\delta_{1}/\Sigma$ & $\gamma_{1}$ & $\gamma_{2}$\\
      \hline
      $0.5$ & $38.217$ & $1878.281$ & $49.352$ & $49.314$ & $-0.038$ & $0.090$ & $-0.419$ & $0.739$ & $-0.758$\\
      \hline
      $1.0$ & $44.593$ & $1904.747$ & $42.873$ & $42.853$ & $-0.020$ & $0.099$& $-0.201$ & $0.331$ & $-1.199$\\
      \hline
      $1.5$ & $44.911$ & $1849.087$ & $41.290$ & $41.299$ & $0.009$ & $0.063$ & $0.145$ & $-0.162$ & $-1.324$\\
      \hline
      $2.0$ & $49.651$ & $1969.613$ & $39.763$ & $39.790$ & $0.027$ & $0.064$ & $0.421$ & $-0.596$ & $-1.033$\\
      \hline
      $3.0$ & $49.655$ & $1794.555$ & $36.248$ & $36.256$ & $0.008$ & $0.037$ & $0.222$ & $-0.261$ & $-0.941$\\
      \hline
      $\textrm{Planck}$ & $54.561$ & $1933.400$ & $35.444$ & $35.318$ & $-0.126$ & $0.564$ & $-0.224$ & $1.435$ & $2.028$\\
      \hline
      $\infty$ & $59.358$ & $2020.043$ & $34.130$ & $34.130$ & $1.3\times10^{-4}$ & $0.001$ & $0.09$ & $-0.248$ & $-0.031$\\
      \hline
      \caption{
        This table shows the values of $\langle\sigma_{0}\rangle$, $\langle\sigma_{1}\rangle$ and $\langle\rho\rangle$ obtained by mask--sweeping.
        Each line correspond from top to bottom to each of the $\rho-$carpets shown in figures \ref{fig:fig5}, \ref{fig:fig6}, \ref{fig:fig7}, \ref{fig:fig8}, \ref{fig:fig9} for the tori, in \ref{fig:fig10} for the Planck map and \ref{fig:fig4} for the spatially infinite $\Lambda$CDM model. The table also lists the median of $\rho$, $\delta_{1}$ := $median$ - $\langle\rho\rangle$, the standard deviation
          $\Sigma$, the ratio $\delta_{1}$/$\Sigma$ and the skewness
        $\gamma_{1}$ := $\left< (\rho - \left< \rho \right>)^{3}\right>/\Sigma^{3}$ and the kurtosis
        $\gamma_{2}$ := [$\left< (\rho - \left< \rho \right>)^{4}\right>/\Sigma^{4}$] - 3 for 768 different orientations of the U73 mask.
        For each of these 768 mask orientations, averages and statistics of $\sigma_{0}$, $\sigma_{1}$ and $\rho$ are made over 100$\,$000 masked maps 
        of the five different tori and the infinite $\Lambda$CDM model and over the 3 by-product maps (NILC, SMICA and SMICA no SZ) of the CMB. }
      \label{tab:table2}
    \end{longtable}
  \end{center}
\end{footnotesize}

\begin{figure}[!htb]
\includegraphics[height=0.38\textheight,width=1.10\textwidth, angle=0]{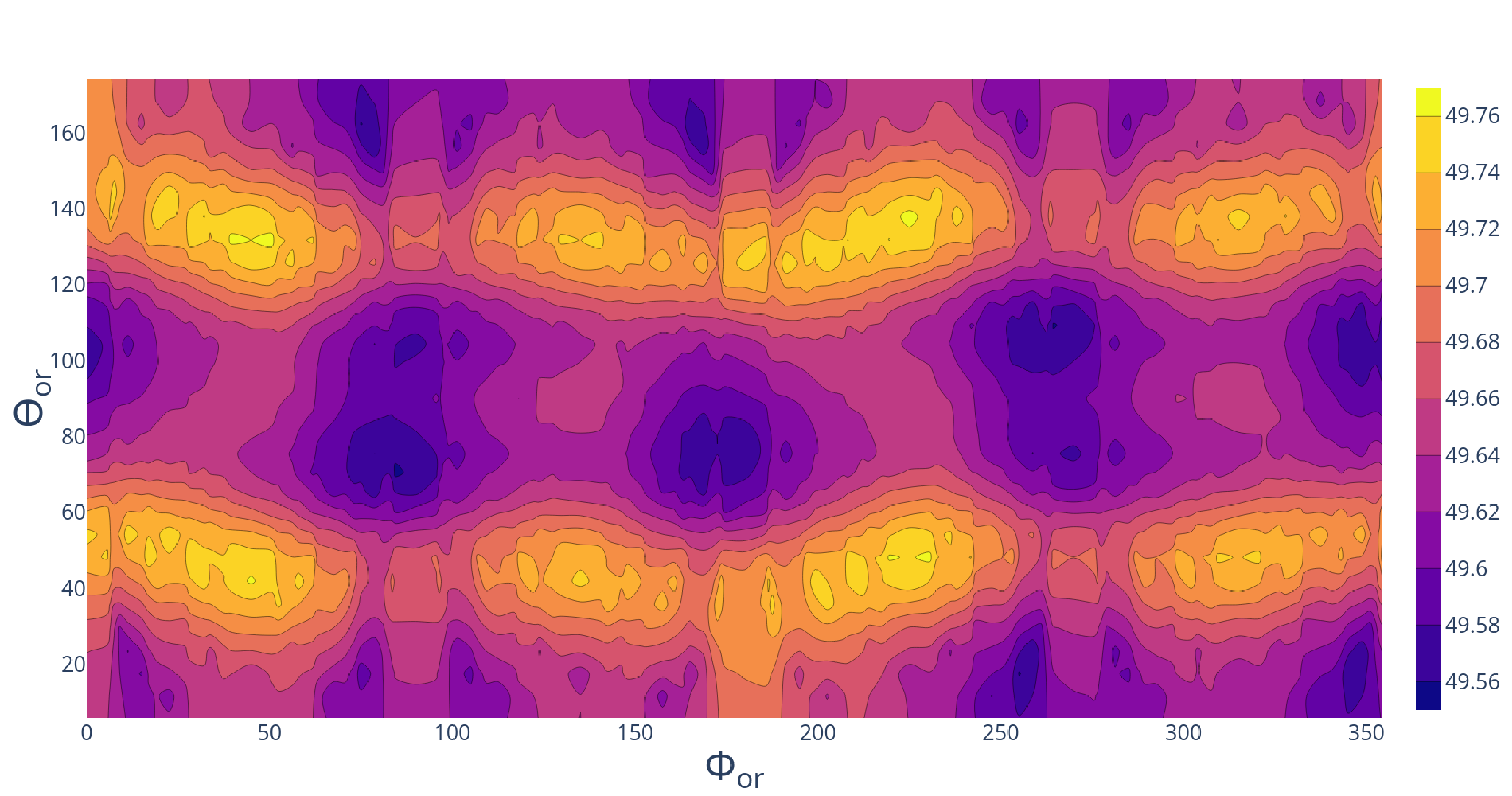}
\caption{This figure shows the $\sigma_{0}$-carpet for the $\Lambda$CDM model with cubic toroidal topology at $L$=3.0.
  This $\sigma_{0}$-carpet is calculated by 768 rotations of the U73 mask over 100$\,$000 maps. The vertical bar gives the scale of $\sigma_{0}$ values in unit of $\muK$.
  These structures are mainly reciproc to those of figure \ref{fig:fig9} because of equation \ref{equ:rho}. 
}
\label{fig:fig11}
\end{figure}

\begin{figure}[!htb]
\includegraphics[height=0.38\textheight,width=1.10\textwidth, angle=0]{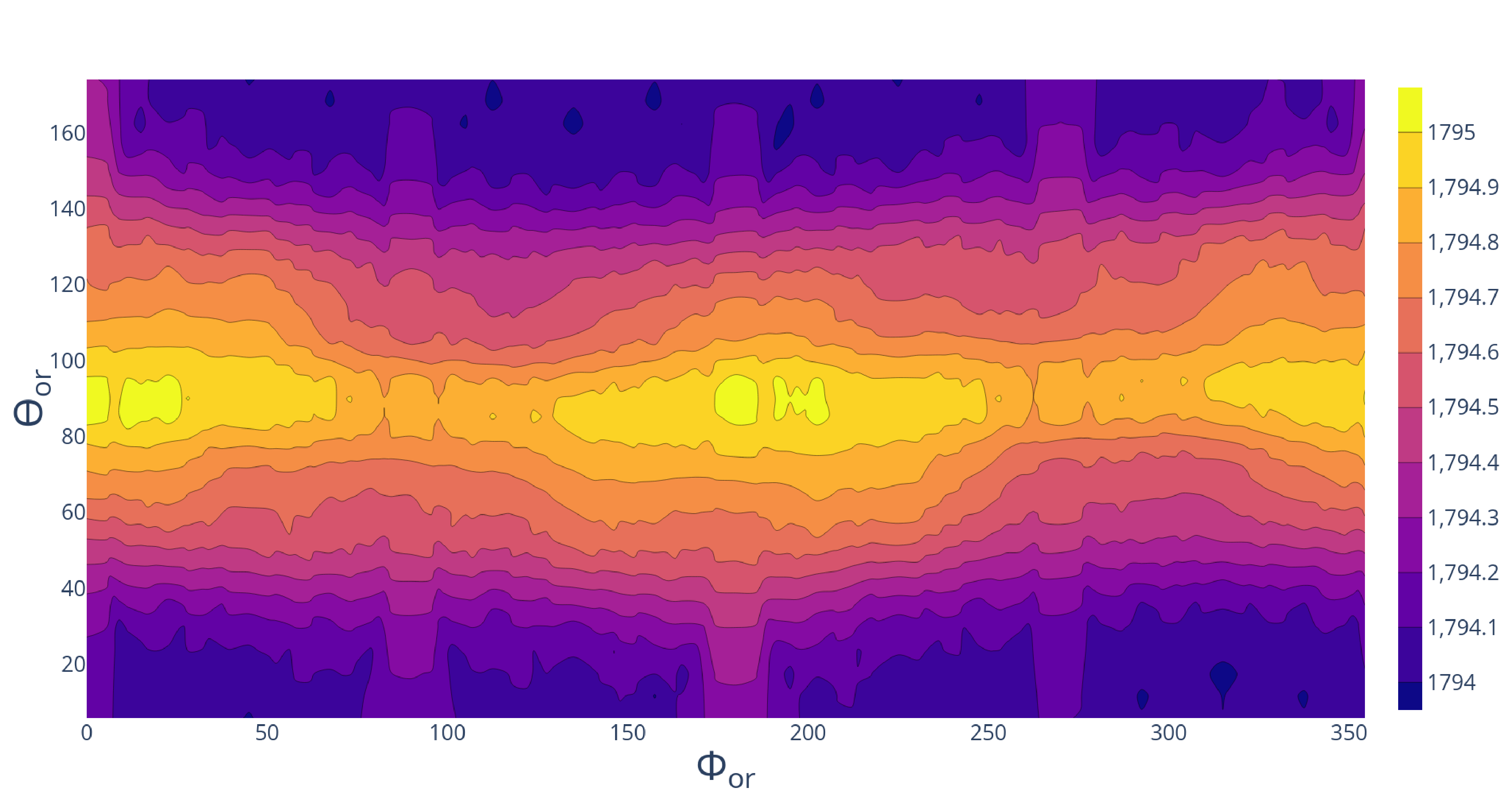}
\caption{This figure shows the $\sigma_{1}$-carpet for the $\Lambda$CDM model with cubic toroidal topology at $L$=3.0.
  This $\sigma_{1}$-carpet is calculated by 768 rotations of the U73 mask over 100$\,$000 maps. The vertical bar gives the scale of $\sigma_{1}$ values in unit of $\muK$.}
\label{fig:fig12}
\end{figure}

Table \ref{tab:table2} gives the statistical results calculated for 768 different mask orientations over the CMB map ensembles of the $3-$torus models
  ($L$=0.5, 1.0, 1.5, 2.0, 3.0) and of the infinite $\Lambda$CDM model in comparison with the unique CMB map. $\sigma_{0}-$ and $\rho-$carpets of
  the $3-$torus model show pronounced periodic patterns with minima and maxima of the torus unveiled by the mask--sweeping
  process over large map ensembles. A single torus map or a too small map ensemble of the torus unveil no periodic pattern. Table 2 in \cite{aurich10}
  (see page 16) should not be
  compared to Table \ref{tab:table2} shown in the present work. In the first case we studied 100$\,$000 maps for a unique mask orientation arriving to a high level
  of Gaussianity. In the present study the mask--sweeping process unveils the violation of statistical isotropy emerging from the background signal in 100$\,$000 maps.
  Thus the mask--sweeping process reveals the underlying morphology specific to the multiply connected topology.
  Conversely, for each of the 768 individual mask orientations, each of the large ensembles of torus maps stay highly Gaussian in the sense we assessed in our previous work
  \cite{aurich10}. 
An interesting outcome of this study is also that the two constituents of $\rho$, $\sigma_{0}$ and $\sigma_{1}$ are both sensitive in a different way to the anisotropy
of the temperature maps.
$\sigma_{0}$ is sensitive to the circle-in-the-sky (CITS) signal specific of each cubic torus size (see figure \ref{fig:fig11}). Indeed the $\sigma_{0}-$carpets
  outline the map of CMB anisotropy while the 
$\sigma_{1}-$carpets have the interesting property to show a rather similar structured pattern for each torus size but they do not correspond to the CITS layout
(see figure \ref{fig:fig12}).
Furthermore, $\sigma_{0}-$, $\sigma_{1}-$ and $\rho-$carpets have in common to show no periodic structure at all for the infinite $\Lambda$CDM model. The CMB in the infinite
$\Lambda$CDM model is statistically isotropic, homogeneous and is a Gaussian random field (see e.g. \cite{buchert6}) for which it is interesting to assess that the
$x-$carpet statistics confirm the absence of anisotropy and periodic feature ($x$ standing for $\sigma_{0}$, $\sigma_{1}$ or $\rho$). 
  
The dissimilarity between the $\rho-$carpets of the $\Lambda$CDM model and the tori is statistically quantifiable with $\delta_{1}$ and the ratio
$\delta_{1}$/$\Sigma$ shown in Table \ref{tab:table2} which characterise the Gaussianity of the $\rho-$carpets. $|\delta_{1}|$=1.3 $ \cdot $ 10$^{-4}$ and $|\delta_{1}$/$\Sigma|$=
0.09 are small for the infinite $\Lambda$CDM model which signal a high Gaussianity. $|\delta_{1}|$ (between 0.008 and 0.038) and $|\delta_{1}$/$\Sigma|$ (between 0.145 and 0.421)
are bigger for the tori. Thus compared to the infinite $\Lambda$CDM model, for the tori $|\delta_{1}|$ is 60 to 300 times larger and $|\delta_{1}$/$\Sigma|$ is 1.6 to 4.7 times larger. 

The $\rho-$carpets of the torus models calculated over sufficiently large CMB maps ensembles (such as the number of 100$\,$000 maps used) are not random fields. We have
calculated the histograms (not shown in the present work) of the $\rho-$carpets. In the case of the torus, the histograms of large ensembles display a strong deviation from
Gaussianity (quantified in Table \ref{tab:table2}). The histogram in the case of the infinite $\Lambda$CDM model corresponds to a much more Gaussian random field as confirmed in Table
\ref{tab:table2}. Indeed it is interesting to observe in the $\rho-$carpets that the background Gaussianity dominant for very small ensemble sizes of a torus gradually
vanishes for larger and larger ensemble sizes putting into evidence the periodic patterns which become dominant. No such periodic structures are expected on the CMB of the
infinite Universe and actually we observe none whatever the choosen size of the ensemble.   

\section{The linear relation between $\rho$ and $L$}
\label{sec:linear}

The exploration of the model maps over 768 different orientations of the U73 mask, regularly distributed on the whole sphere, provides the opportunity to test the linear relation
$L$=f($\langle\rho\rangle$) presented in \cite{aurich10}. We verify here the validity of the linear relation connecting the ratio
$\langle\rho\rangle$=$\sqrt{\langle\frac{\sigma_{1}^{2}}{\sigma_{0}^{2}}\rangle}$ to the side--length $L$ of the cubic $3-$torus $L$$\simeq$a$\langle\rho\rangle+b$.

\begin{figure}[!htb]
\includegraphics[height=0.7\textheight,width=0.50\textwidth, angle=270]{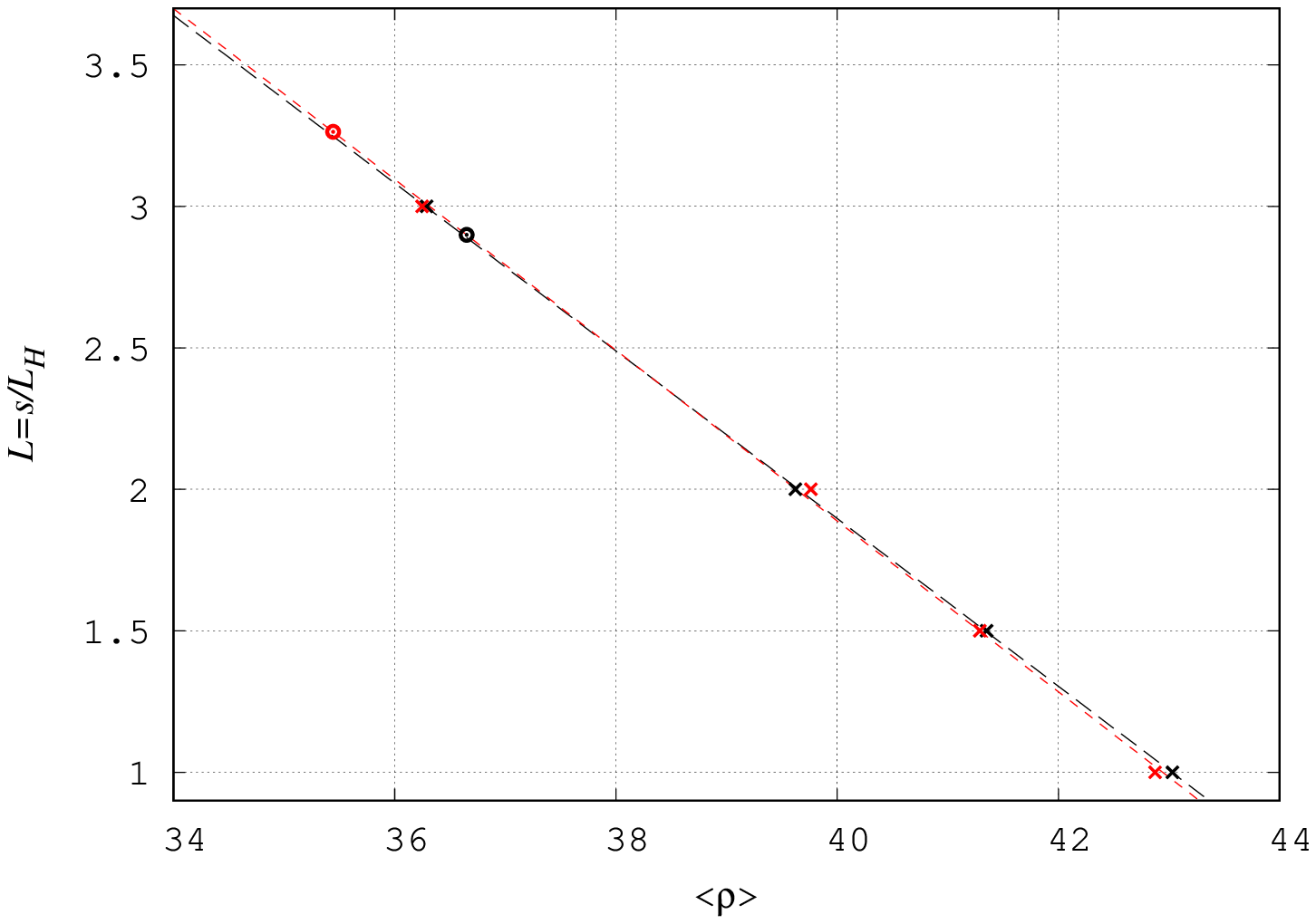}
\caption{Comparison of the two linear regressions $L$$\simeq$a$\langle\rho\rangle$+b with mask $U73$ of the points ($\rho$,$L$), $L$=(1.0,1.5,2.0,3.0), in black for the mask orientation
  by default and in red for the mask sweeping over 768 orientations. The two rings in black and red give an estimate of the cubic $3-$torus size from the average $\rho$ values
  of the observed CMB radiation map. The black ring ($\langle\rho\rangle$=36.650, $L$$\simeq$2.90) is obtained from Planck by-products I intensity maps of the pipelines NILC, SEVEM, SMICA and
  Commander-Ruler. The red ring ($\langle\rho\rangle$=35.444, $L$$\simeq$3.26) is obtained from Planck by-products inpainted I intensity maps of the pipelines NILC, SMICA and SMICA-no SZ.}
\label{fig:fig13}
\end{figure}

Figure \ref{fig:fig13} shows the points $(\rho,L)$ computed with the single mask orientation used in \cite{aurich10} (black crosses) as well as those obtained from the additional average
over the 768 different mask orientations (red crosses). There is almost no difference visible. As a consequence, the least square fitting of $L$$\simeq$a$\langle\rho\rangle+b$
is also almost indistinguishable for both cases, which are also plotted in figure \ref{fig:fig13}. 

\begin{figure}[!htb]
\includegraphics[height=0.7\textheight,width=0.50\textwidth, angle=270]{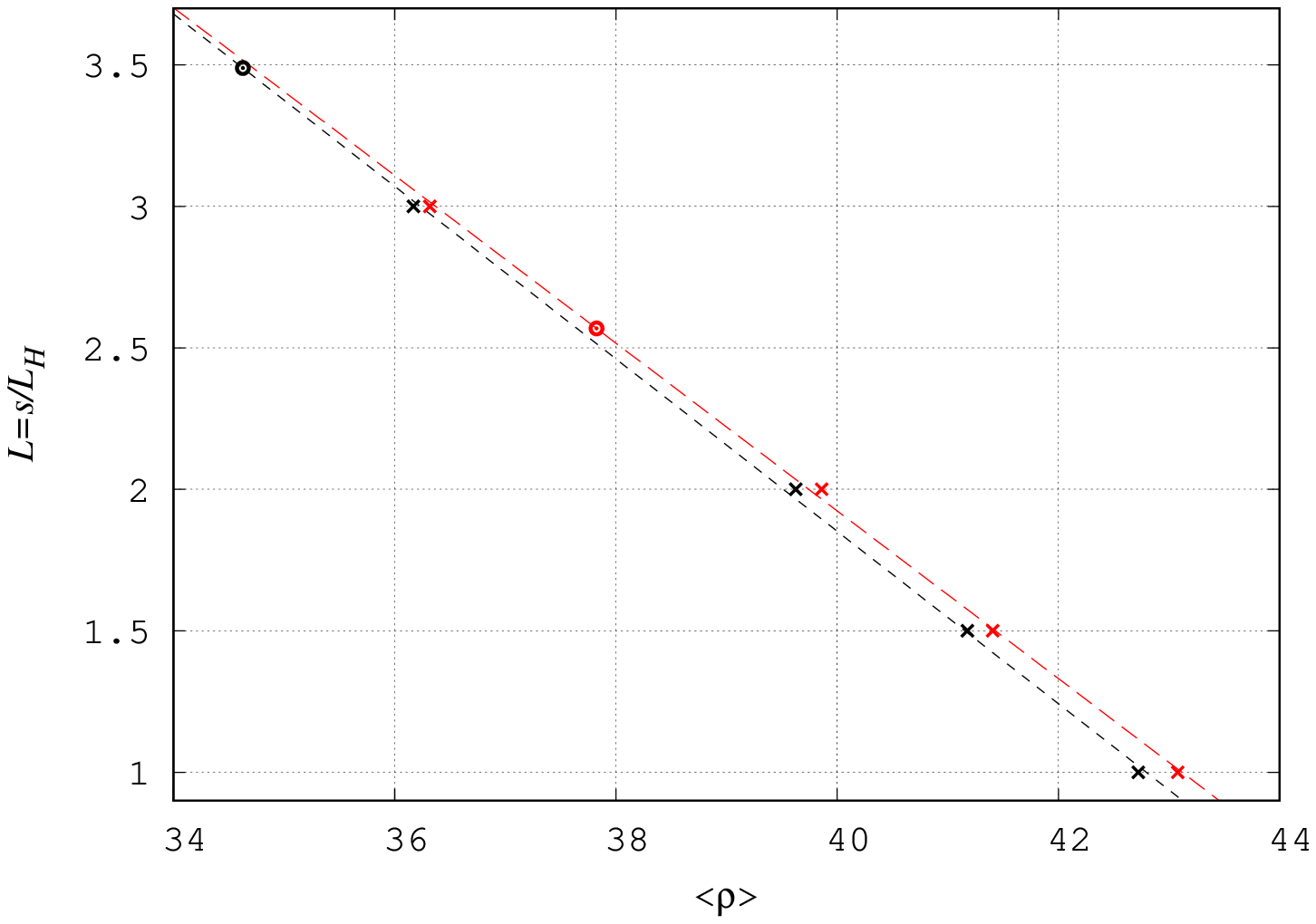}
\caption{This figure shows the two linear regressions for the absolute minimum (in black) and for the absolute maximum (in red) of $\rho$ obtained by mask--sweeping over
  the four tori at $L$=(1.0,1.5,2.0,3.0). The $\rho$ values, 34.628 (black ring) and 37.825 (red ring) for the $Planck$ map yield a $3-$torus size 3.49$>L>$2.57.}
\label{fig:fig14}
\end{figure}

The fit of the linear regressions in figure \ref{fig:fig13} correspond to a $\chi^{2}$
smaller than 10$^{-6}$ for the unique mask orientation while $\chi^{2}$ is barely above 10$^{-5}$ with mask--sweeping (see also in figure \ref{fig:fig14}). Given the $\rho$ values
(36.650 and 35.444) obtained over the $Planck$ pipelines NILC, SMICA and SMICA-no SZ of the CMB radiation with mask, one is led to the conclusion that our
Universe could have a topology equivalent to the one of a cubic $3-$torus with side--length 2.90$<L<$3.26.

\smallskip

As a further test for the robustness of the linear relation stated in \cite{aurich10}, we determine from each $\rho-$carpet the maximum and the minimum. From these extreme values,
the linear regression is determined as above. The result is shown in figure \ref{fig:fig14} together with the corresponding extremal $\rho$ values of 37.825 and 34.628 obtained from the
Planck map. Figure \ref{fig:fig14} reveals that also ``worst case'' mask orientations follow the original linear relation found with the single mask orientation used in
\cite{aurich10}. The violation of statistical isotropy is almost negligible in the case of the cubic $3-$torus topology and emerge only in the analysis of sufficiently large ensemble
sizes. 

\bigskip

\section{Conclusion}
\label{conclusion}

The present work proves that very faint persistent geometrical and morphological features of the CMB temperature anisotropy maps, invisible over a too small map ensemble, can be
revealed by a morpho--statistical process combining calculation of $\sigma_{0}$, $\sigma_{1}$ and the standard deviation of the normalised temperature gradient $\rho$ to a mask--sweeping
over 768 evenly distributed orientations of
the $U73$ $Planck$ mask. This $\rho-$carpet generator analyses 100$\,$000 maps of each given universe model. A specific periodic pattern emerges from thousands of CMB temperature
maps with cubic toroidal topology. By comparison the same number of CMB maps simulated in the infinite $\Lambda$CDM model reveals no periodic structure, as expected, 
but displays a random field varying with the size of the map ensemble. This is due to the residual of averaging a finite ensemble size of a statistically isotropic random field. 

The $\rho-$carpet of the unique CMB map observed by $Planck$ in figure \ref{fig:fig10} doesn't display the 
structures visible in figure \ref{fig:fig9} of the torus at $L$=3.0. But we have seen that the detection method by mask--sweeping reveals persistent structures only for sufficiently
large ensembles of CMB maps.
The mask--sweeping process putting no assumption on the nature of the underlying geometry and anisotropy is in this sense model-independent. This tool reveals the high level
of statistical isotropy and homogeneity of the CMB in the infinite $\Lambda$CDM model.

Table \ref{tab:table2} summarises the statistical values for the different ensembles. 
The values of $|\delta_{1}|$ and more peculiarly $|\delta_{1}$/$\Sigma|$ are a signature of the level of Gaussianity. These two quantities are small for the infinite $\Lambda$CDM
model and thus characterize its high Gaussianity. The large values of $|\delta_{1}|$ and $|\delta_{1}$/$\Sigma|$ for the tori point to the good detection of anisotropy patterns. 

Our new methodology confirms the linear relation $L$$\simeq$a$\langle\rho\rangle+b$
matching to a size of $L$$\sim$3 for our Universe with cubic toroidal topology.
Furthermore, our analysis shows that the minute violations of statistical isotropy in the case of the cubic $3-$torus do not alter the linear relation found in \cite{aurich10}. 

Exploring each of the 100$\,$000 model maps for each of the different universe models using 768 different orientations of the foreground mask represents a computationally heavy task.
Different $n_{\textrm{\scriptsize or}}$ values have been tested over smaller map ensembles. It comes out that the global pattern of periodicity specific to the tori stays the same
whatever the value of $n_{\textrm{\scriptsize or}}$. A compromise has been made in order to outline precisely the layout of spots while keeping the computing time to reasonably small durations.

Our results envision that our Universe might have the topology of a $3-$torus of finite volume. But it has been seen that other multiply connected
topologies, e.g. the Poincar\'e dodecahedron \cite{aurich9} or the Picard manifold \cite{aurich16,aurich17} present a $2-$point correlation function close to the
one of the observable CMB. Therefore it seems that other finite models of our Universe, more complex, more realistic, should also be considered \cite{akrami1}.
Signatures of finite multiply connected manifolds like the $3-$torus remain to be found in the distribution of large structures of the Universe.

\vspace{15pt}
\noindent
{\em Acknowledgements:}
\small{MJF is grateful to Michel Tallon for interesting discussions. We gratefully acknowledge support from the CBPsmn (Centre Blaise Pascal de Simulation et de Mod\'elisation Num\'erique) at
  the \'Ecole Normale Sup\'erieure de Lyon for the computing resources [\href{https://www.ens-lyon.fr/PSMN/doku.php?id=accueil:presentation}{CBPsmn presentation}].
  Regarding software issues and the package HEALPix, we more specifically acknowledge discussions, expertise and advices provided by Cerasela Calugaru and Lo\"is Taullele at CBPsmn and
  \'Eric Hivon at IAP (Institut d'Astrophysique de Paris) [\href{https://www.iap.fr}{IAP}].
  Python 3.12.3 interpreter, the libraries Numpy 1.26.4, Plotly 5.15.0 and the image generator Kaleido 1.2.0 are used during this research. 
This work is also based on observations obtained with $Planck$ [\href{http://www.esa.int/Planck}{Planck}], an ESA science mission with instruments and 
contributions directly funded by ESA Member States, NASA [\href{http://lambda.gsfc.nasa.gov/product/planck/curr/planck_prod_irsa.html}{NASA $\lambda$}], and Canada. Some of the
results in this paper have been derived using the HEALPix package \cite{gorski1}, available at [\href{https://healpix.sourceforge.io/credits.php}{HEALPix}] (update 2025).
The CMB power spectra and the $2-$point correlation function of the infinite $\Lambda$CDM model are calculated from the cosmological parameters using $CAMB$ software written by
Anthony Lewis and Anthony Challinor (update 2024) [\href{http://lambda.gsfc.nasa.gov/toolbox/camb_online.html}{CAMB interface}] from the original Boltzmann codes by Edmund
Bertschinger, Chung-Pei Ma and Paul Bode resumed by Uro\u s Seljak and Matias Zaldarriaga. The $CAMB$ ReadMe is available here [\href{http://camb.info/readme.html}{CAMB readme}]
and $CAMB$ Notes by A Lewis (2025) here [\href{http://cosmologist.info/notes/CAMB.pdf}{CAMB Notes}].}

\section*{ORCID IDs}
\label{orcid}
Ralf Aurich \orcidlink{0000-0002-3206-9462}{\color{ballblue}{ https://orcid.org/0000-0002-3206-9462}}
\newline
Martin J France \orcidlink{0000-0002-2815-6686}{\color{ballblue}{ https://orcid.org/0000-0002-2815-6686}}
\newline
Frank Steiner \orcidlink{0000-0003-2039-0213}{\color{ballblue}{ https://orcid.org/0000-0003-2039-0213}}

\section*{Data availability statement}

We will provide access to our data on reasonable request after a suitable but relatively short embargo period, allowing us to fully exploit their scientific potential. 

\appendix

\section{Orientations of the mask over the $2-$sphere}
\label{app:euler}

An arbitrary orientation of the mask U73 shown in figure \ref{fig:fig3} can be for instance decomposed into a parallel displacement and a tilt.

\smallskip

In virtue of Euler's rotation theorem
in a $3$-dimensional Euclidean space, 

\smallskip

(i) an arbitrary orientation of the mask can be obtained by a unique rotation of angle $\eta$ around a well chosen axis, but in the present case a unique axis
of rotation cannot be arbitrarily chosen,

\smallskip

(ii) a complete exploration of the mask orientations (position plus tilt) on the $2-$sphere requires that for each combination of rotations by two Euler's angles
$\boldsymbol{\vartheta_{\textrm{\scriptsize or}}}$($\Phi_{\textrm{\scriptsize or}}$, $\Theta_{\textrm{\scriptsize or}}$) the space of orientations over a third Euler's angle
$\Xi_{\textrm{\scriptsize or}}$ has to be explored also.

\smallskip

In our implementation, 768 different orientations of the foreground mask are obtained by composition $\boldsymbol{\vartheta_{\textrm{\scriptsize or}}}$
of the two rotations $\Phi_{\textrm{\scriptsize or}}$ and $\Theta_{\textrm{\scriptsize or}}$ using the 2 first angles in the HEALPix routine rotate$\_$alm.
In order to keep reasonable computing durations the third Euler angle of the routine rotate$\_$alm is not explored in our present investigation. 

\section*{References}
\bibliographystyle{iopart-num}

\end{document}